\documentclass{article}

\usepackage{arxiv}
\renewcommand{\headeright}{}
\renewcommand{\undertitle}{}
\usepackage[utf8]{inputenc}
\usepackage[T1]{fontenc}
\usepackage{microtype}
\usepackage{amsmath,amssymb,mathtools}
\usepackage{xcolor,colortbl}
\usepackage{graphicx}
\graphicspath{{figures/}}
\usepackage{flafter,placeins}
\usepackage{booktabs,array,tabularx}
\usepackage{fvextra}
\usepackage[authoryear,round]{natbib}
\setcitestyle{citesep={;},aysep={,},yysep={;}}
\usepackage{url}
\usepackage[hidelinks,breaklinks=true]{hyperref}

\definecolor{lightblue}{RGB}{70,140,210}
\newsavebox{\resulttablebox}
\newcommand{\modelname}{MedSonor}
\newcommand{\baseModel}{Qwen2.5-Omni-7B}
\newcommand{\judgeModel}{Qwen3-4B-Instruct}
\DefineVerbatimEnvironment{promptcode}{Verbatim}{
  fontsize=\scriptsize,
  frame=single,
  framesep=3pt,
  rulecolor=\color{black!35},
  breaklines=true
}

\title{\modelname: Learning to Use\\Clinical Audio in Medical LLMs}
\renewcommand{\shorttitle}{MedSonor: Learning to Use Clinical Audio in Medical LLMs}
\author{%
  Geyang Yu\textsuperscript{1}\\
  \normalfont The Chinese University\\
  \normalfont of Hong Kong, Shenzhen\\
  \small\href{mailto:225085005@link.cuhk.edu.cn}{\texttt{225085005@link.cuhk.edu.cn}}
  \And
  Junying Chen\textsuperscript{2}\\
  \normalfont The Chinese University\\
  \normalfont of Hong Kong, Shenzhen\\
  \small\href{mailto:junying.chen.cs@gmail.com}{\texttt{junying.chen.cs@gmail.com}}
  \And
  Benyou Wang\thanks{Corresponding author.}\\
  \normalfont The Chinese University\\
  \normalfont of Hong Kong, Shenzhen\\
  \small\href{mailto:wangbenyou@cuhk.edu.cn}{\texttt{wangbenyou@cuhk.edu.cn}}
}

\date{}
\hypersetup{
  pdftitle={MedSonor: Learning to Use Clinical Audio in Medical LLMs},
  pdfauthor={Geyang Yu, Junying Chen, Benyou Wang},
  pdfsubject={Medical audio-language models and rubric-guided reinforcement learning}
}

\begin{document}

\maketitle

\begin{abstract}
Clinical audio conveys medical information that transcripts may not preserve,
but using it in medical responses requires interpreting it alongside clinical
requests and dialogue context. Existing medical audio supervision largely
targets classification or recognition, while general audio-language models
lack specialized medical knowledge and direct feedback for open-ended
responses. We present \modelname{}, a medical dialogue model integrating
medical knowledge, spoken dialogue, and clinical audio understanding. We pair
authentic medical audio with clinical requests and combine these
audio-integrated dialogues with collected medical audio and synthesized
spoken dialogues for supervised fine-tuning, followed by two stages of
rubric-guided reinforcement learning. We introduce MedVoice-Dial and
MedVoice-AID to evaluate responses within supplied multi-turn medical
contexts and to requests involving clinical audio, respectively.
Across seven medical benchmarks, \modelname{} outperforms the evaluated
general-purpose open-weight audio-language baselines and ranks highest
in aggregate physician preference evaluation, while broadly retaining
general speech capabilities.
\end{abstract}

\section{Introduction}
\label{sec:introduction}
Clinical assessment draws on patient reports and acoustic information that
transcripts may omit. Respiratory and cardiac recordings support sound and
murmur analysis~\citep{rocha2019respiratory,oliveira2022circor}, while voice
research spans laryngeal, neurological, and mood-related conditions~\citep{bensoussan2024voice}.
While large language models (LLMs) show promise in medical reasoning and
consultation~\citep{medpalm,meditron,amie}, medical voice assistants must also
interpret these acoustic cues alongside the clinical request and dialogue context.

However, existing audio datasets do not fully reflect real-world clinical
tasks, limiting their ability to supervise useful medical responses.
Bridging this gap also requires adapting medical knowledge to acoustic
inputs. Further improving these responses remains difficult without
task-specific feedback for medical audio dialogue.

We present \modelname{}, an audio-conditioned medical dialogue model adapted
from Qwen2.5-Omni-7B. We address these challenges through three components.
\textbf{(1) Contextualized task construction:}
We collect and standardize medical audio, synthesize spoken multi-turn dialogues across clinical specialties and user roles, and construct audio-integrated dialogues that pair authentic audio with clinical requests, connecting existing audio resources to medical response tasks.
\textbf{(2) Adapting medical knowledge to audio-based tasks:}
We jointly fine-tune the backbone on these complementary sources to develop medical knowledge, medical audio understanding, and context-sensitive response generation.
\textbf{(3) Rubric-guided response optimization:}
We define instance-specific rubrics covering required content, dialogue constraints, and prohibited errors, and train local item-level judges to provide feedback. These signals guide two successive reinforcement learning stages for medical dialogue and audio-integrated requests. Severity-based reward caps prevent high overall rewards when the judges detect major or critical violations.

We evaluate \modelname{} using MedVoice-Dial for responses within supplied multi-turn medical contexts and MedVoice-AID for clinical requests containing medical audio, together with five public or speech-adapted medical benchmarks. \modelname{} ranks first among the evaluated general-purpose open-weight audio-language baselines on all seven benchmarks and achieves the highest aggregate physician preference ranking in a study of 30 MedVoice-AID cases. Removing audio-integrated dialogue supervision reduces the MedVoice-AID score from 0.7900 to 0.5957, supporting the value of explicitly pairing medical audio with clinical requests. Six general speech evaluations further show broad retention of the backbone's capabilities. Together, these findings support our approach to adapting general audio-language models to answer clinical questions.

Our contributions are: 1) We introduce \modelname{}, combining multimodal medical SFT with two-stage RL guided by instance-specific rubrics and severity-based reward caps. 2) We construct medical audio and dialogue resources and introduce MedVoice-Dial and MedVoice-AID to evaluate medical dialogue responses informed by consultation history and responses to clinical audio. 3) \modelname{} leads the evaluated general-purpose open-weight audio-language baselines across seven medical benchmarks and in aggregate physician preference on 30 cases. Ablations support explicit audio--request pairing, with general speech capabilities broadly retained.

\section{Background}
\label{sec:background}

\subsection{Clinical Role of Medical Audio}
\label{sec:clinical_audio}

Medical audio serves two complementary roles: conveying spoken symptoms,
histories, and requests, and providing acoustic evidence that transcripts may
not preserve.
Respiratory and cardiac sounds support auscultation, while voice quality,
articulation, and prosody inform research on laryngeal function, neurological
status, and emotional state~\citep{rocha2019respiratory,oliveira2022circor,bensoussan2024voice}.
Pain-related vocalizations can complement patient reports, while infant cries
provide cues when verbal self-report is unavailable~\citep{dao2025tame,bellieni2004cry}.
Table~\ref{tab:clinical_audio_roles} summarizes representative categories and
their clinical relevance.
These roles make audio both a communication channel and a source of clinical
evidence: effective assistance must understand what the user asks and determine
what the available sounds can support.

\begin{table}[!htbp]
\centering
\footnotesize
\caption{Clinical roles of medical audio, spanning physiological sounds,
spoken information, and vocal cues. Acoustic findings require clinical context
and do not establish diagnoses alone.}
\label{tab:clinical_audio_roles}
\setlength{\tabcolsep}{4pt}
\renewcommand{\arraystretch}{1.05}
\begin{tabularx}{\linewidth}{
    @{}
    >{\raggedright\arraybackslash}p{0.215\linewidth}
    >{\raggedright\arraybackslash}p{0.335\linewidth}
    >{\raggedright\arraybackslash}X
    @{}
}
\toprule
\textbf{Audio category}
& \textbf{Key cues}
& \textbf{Clinical relevance} \\
\midrule
Lung sounds
& Wheezes, crackles, breath patterns
& Respiratory auscultation. \\
Snoring
& Spectral and temporal patterns
& Upper-airway obstruction research. \\
Cardiac audio
& Heart sounds and murmurs
& Murmur detection and characterization. \\
Cough
& Cough events and acoustics
& Symptom assessment and screening research. \\
\midrule
Medical speech
& Symptoms, history, concerns
& History taking and clinical reasoning. \\
Spoken intent
& Questions and requests
& Consultation goals and follow-up. \\
\midrule
Voice disorders
& Voice quality and articulation
& Voice and motor-speech assessment. \\
Pain-related speech
& Speech changes and vocal cues
& Complementing pain reports in research. \\
Infant cries
& Pitch, intensity, timing
& Pain assessment without self-report. \\
Affective speech
& Prosody and emotion
& Mood-related research and communication. \\
\bottomrule
\end{tabularx}
\end{table}

\subsection{From Sound Interpretation to Clinical Responses}
\label{sec:background_challenges}

Representative models emphasize different aspects of medical audio and
interaction (Table~\ref{tab:related_work_summary}).
HeAR focuses on health-acoustic representations, AMIE on text-based
consultation, and Qwen2.5-Omni on general audio understanding and dialogue.
SpeechMedAssist and StethoLM already connect medical sounds with spoken
consultation or instruction-driven clinical tasks.
Building on these directions, our focus is on supervision that connects
acoustic evidence with responses across clinical requests and interaction
contexts.

Recognizing a sound does not by itself determine how to answer a medical
question.
The same recording may prompt a request for a sound description, an explanation
of its relevance, or guidance on what information is needed next.
The appropriate response therefore depends on the recording, the current
request, and any supplied dialogue history.
Conversely, a fluent medical answer may misinterpret the sound or draw
conclusions that the evidence cannot support.
The learning objective must consequently extend beyond acoustic recognition
to \emph{using acoustic evidence in context-sensitive medical responses}.

\begin{table}[!htbp]
\centering
\footnotesize
\caption{Reported scope of representative models. Entries describe task
coverage, not equivalent performance or clinical validation.}
\label{tab:related_work_summary}
\setlength{\tabcolsep}{5pt}
\renewcommand{\arraystretch}{1.08}
\begin{tabularx}{\linewidth}{
    @{}
    >{\raggedright\arraybackslash}p{0.32\linewidth}
    >{\raggedright\arraybackslash}X
    >{\raggedright\arraybackslash}X
    @{}
}
\toprule
\textbf{Work}
& \textbf{Audio modeling}
& \textbf{Dialogue and response tasks} \\
\midrule
HeAR~\citep{hear2024}
& Health-acoustic representations
& Consultation outside reported scope \\

Qwen2.5-Omni~\citep{qwen25omni}
& General audio understanding
& General-purpose dialogue \\

AMIE~\citep{amie}
& Clinical sound interpretation outside reported scope
& Text-based clinical consultation \\

SpeechMedAssist~\citep{speechmedassist}
& Cough and other vocal cues
& Spoken medical consultation \\

StethoLM~\citep{stetholm}
& Cardiopulmonary sound interpretation
& Instruction-driven clinical tasks \\
\bottomrule
\end{tabularx}
\end{table}

\noindent\textbf{Motivation for \modelname{}}
Effective medical audio dialogue requires not only recognizing sounds, but
learning how acoustic evidence should inform a response to a clinical request.
Sound-centered datasets provide limited supervision for this connection,
while spoken dialogues synthesized from text do not by themselves teach
interpretation of authentic clinical recordings.
Moreover, an accurate sound description may fail to address the clinical
question, while a medically plausible answer may contradict the recording
or dialogue history.
These gaps motivate \modelname{}, which combines collected medical audio,
synthesized spoken dialogues, and audio-integrated dialogues to connect
acoustic understanding with contextual medical responses.
Beyond these demonstrations, two-stage rubric-guided RL provides
instance-specific feedback on acoustic interpretation, clinical
appropriateness, and dialogue consistency.
The rubrics distinguish required content from prohibited errors, with
severity-based reward caps preventing high rewards when major or critical
violations are detected.
Together, this supervision and feedback design targets both
\emph{understanding clinical audio} and \emph{using it appropriately in
medical responses}, as detailed in Section~\ref{sec:method}.

\section{Method}
\label{sec:method}

\subsection{Overview}
\label{sec:method_overview}

\modelname{} receives user audio and, when provided by the task, dialogue
history and transcript context.
Its Qwen2.5-Omni backbone supports text and speech generation through a
Thinker--Talker architecture: the Thinker produces text, and the Talker
conditions on its hidden representations and text to generate speech
tokens~\citep{qwen25omni}. For consistent evaluation of response content,
we use the Thinker's text outputs.

\subsection{Data Construction}
\label{sec:task_formulation}

\begin{figure}[!htbp]
\centering
\includegraphics[width=1\linewidth]{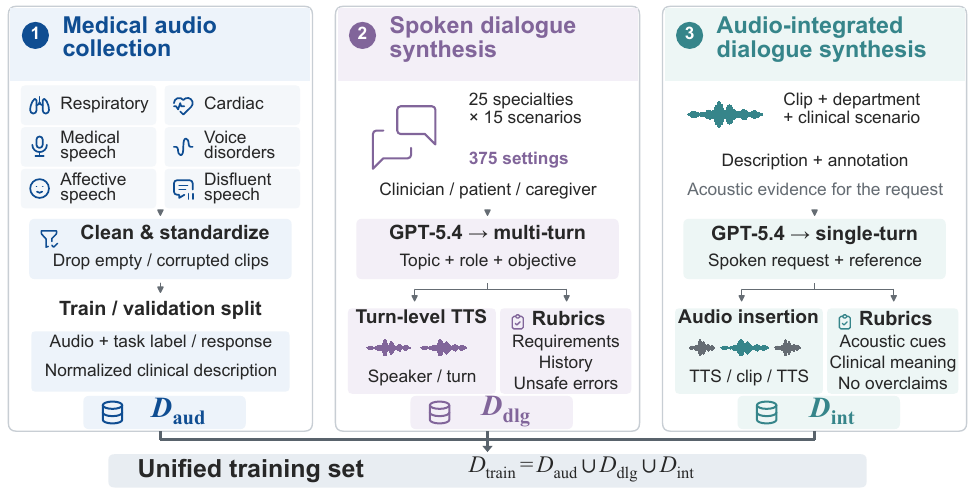}
\caption{Data construction: standardized medical audio, synthesized spoken
multi-turn dialogues with turn-level rubrics, and single-turn dialogues
incorporating authentic medical audio.}
\label{fig:data_construction_pipeline}
\end{figure}

We construct three complementary sources that connect medical audio
understanding with clinical response tasks, together with rubrics for
training feedback (Figure~\ref{fig:data_construction_pipeline}).

\noindent  \textbf{Collected medical audio.}
\label{sec:medical_audio_collection}
To establish medical audio understanding, we curate physiological sounds and
clinically relevant speech, including respiratory and cardiac sounds, medical
speech with transcripts, recordings of emotional or disfluent speech, and
recordings of voices affected by voice disorders. Examples
contain audio, a task category, an available label or response, and a normalized
clinical description. This stream includes 349 waveforms reconstructed from
Bridge2AI-Voice v3.0.0 spectrograms, rather than original recordings
(Appendix~\ref{app:bridge2ai_reconstruction}). We remove empty or corrupted audio clips, standardize
audio and metadata, and construct training and validation splits
(Appendix~\ref{app:audio_details}).

\noindent  \textbf{Synthesized spoken dialogue.}
\label{sec:medical_dialogue_synthesis}
To support contextual medical interaction, we combine 25 clinical specialties with 15
scenarios, forming 375 settings with specified topics, user roles, and
objectives. GPT-5.4 generates multi-turn dialogues spanning professional,
patient, and caregiver interactions. Utterance-level TTS preserves speaker
and turn alignment. Scenario definitions and corpus statistics appear in
Appendices~\ref{app:data_synthesis_details} and~\ref{app:data_details}.

\noindent  \textbf{Audio-integrated dialogue.}
\label{sec:audio_integrated_dialogue_synthesis}
To connect audio with clinical requests, we assign each audio clip a
department using its provenance and clinical category, then pair it with the
same 15 scenarios. GPT-5.4 uses the audio clip's description and annotation to
generate a single-turn request and reference response. We synthesize the
request with TTS and insert the authentic audio clip at the specified position,
preserving segment boundaries and order.

\noindent  \textbf{Instance-specific rubrics.}
GPT-5.4 generates rubrics from the request, reference response, and relevant
context: preceding user and reference-assistant turns for spoken dialogue,
or the audio clip's description for audio-integrated requests. The criteria
specify required content, contextual constraints, and prohibited errors,
providing item-level feedback for reinforcement learning.

\subsection{Training}
\label{sec:training_pipeline}

\begin{figure}[!htbp]
\centering
\includegraphics[width=1\linewidth]{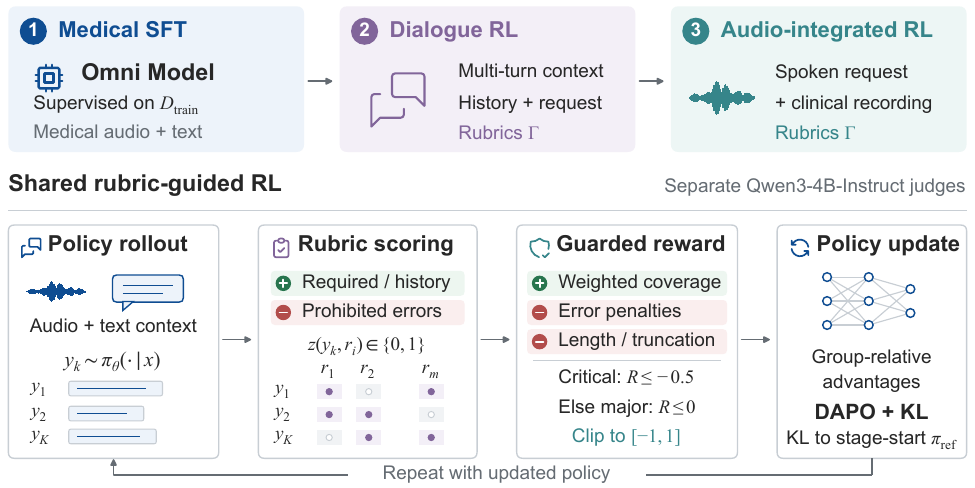}
\caption{Post-training overview. Medical SFT initializes the policy; successive
medical dialogue RL and audio-integrated dialogue RL stages optimize severity-guarded rubric rewards.}
\label{fig:method_overview}
\end{figure}

As shown in Figure~\ref{fig:method_overview}, training proceeds from medical
SFT to medical dialogue RL and then audio-integrated dialogue RL, with both
RL stages guided by severity-guarded rubric rewards.

\noindent  \textbf{Multimodal medical SFT.}
\label{sec:sft_stage}
Medical SFT adapts the general-purpose backbone to medical content and response requirements through collected medical audio, synthesized spoken dialogues, and audio-integrated dialogues. Inputs may contain text, medical audio, spoken requests, and
dialogue history. Configuration details are reported in
Appendix~\ref{app:training_details}.

\noindent  \textbf{Item-level judges.}
\label{sec:local_verifier}
\label{sec:rubric_formulation}
Instance-specific rubrics translate the requirements of each medical request into item-level feedback for optimizing open-ended responses. We train separate local Qwen3-4B-Instruct-2507 (\judgeModel{}) judges for dialogue and
audio-integrated responses. A rubric comprises positive requirements
$\Gamma^{+}$, prohibited errors $\Gamma^{-}$, and history-derived constraints
$\Gamma^{h}$. GPT-5.4 labels whether candidate responses satisfy positive or
history items and trigger negative items. These binary labels supervise
$z(H,u,y,r)\in\{0,1\}$ for response $y$ and rubric item $r$; we omit the
conditioning context below for brevity. Training and validation details for
the local judges appear in Appendix~\ref{app:local_judge_details}.

\noindent  \textbf{Severity-guarded reward.}
\label{sec:guarded_reward}
\label{sec:audio_guarded_reward}
Both RL stages use normalized weighted coverage of positive and history items,
minus negative-item penalties and length/truncation penalties:
\begin{equation}
    \widetilde{R}_{\Gamma}(y)
    =
    \frac{\sum_{r_i\in\Gamma^{+}\cup\Gamma^{h}}w_i z(y,r_i)}
         {\sum_{r_i\in\Gamma^{+}\cup\Gamma^{h}}w_i+\epsilon}
    -
    \frac{\sum_{r_j\in\Gamma^{-}}|w_j|z(y,r_j)}
         {\sum_{r_j\in\Gamma^{-}}|w_j|+\epsilon}
    -P_{\mathrm{len}}(y)-P_{\mathrm{trunc}}(y).
\end{equation}
Here $w_i$ is an item weight and $\epsilon$ stabilizes normalization.
The terms $P_{\mathrm{len}}(y)$ and $P_{\mathrm{trunc}}(y)$ penalize excessive
response length and truncated responses, respectively; their definitions and
settings are provided in Appendix~\ref{app:reward_details}.
The final reward $R_{\Gamma}(y)$ caps this score at zero for a recognized major
violation and at $-0.5$ for a critical violation, then clips it to $[-1,1]$.

\noindent  \textbf{Sequential rubric-guided RL.}
\label{sec:rl_stage}
\label{sec:filter_train}
Medical dialogue RL trains on multi-turn instances $(H,u,\Gamma)$ from the synthesized dialogue
corpus. For each input, the policy samples a group of responses, the dialogue
judge supplies rubric rewards, and DAPO~\citep{dapo} updates the policy using
group-relative advantages. Both stages optimize reward with KL regularization
against the policy that initializes the stage:
\begin{equation}
    \max_{\theta}\;
    \mathbb{E}_{x\sim\mathcal{D}_{\mathrm{RL}},\,
    y\sim\pi_{\theta}(\cdot\mid x)}
    \left[R_{\Gamma}(y)\right]
    -\beta\,D_{\mathrm{KL}}\!\left(
    \pi_{\theta}(\cdot\mid x)\,\|\,\pi_{\mathrm{ref}}(\cdot\mid x)
    \right).
\end{equation}
\label{sec:audio_dialogue_rl}
Audio-integrated dialogue RL applies the same optimization to spoken requests combined with clinical audio recordings and their associated rubrics,
using the separate audio-integrated judge. Data filtering and
optimization settings are provided in Appendices~\ref{app:data_details} and~\ref{app:training_details}.
\section{Experiments}
\label{sec:experiments}

\subsection{Evaluation Protocol}
\label{sec:evaluation_protocol}

Medical response evaluation covers LiveMedBench Clean-TTS~\citep{livemedbench},
MMedFD-Lite~\citep{mmedfd}, an ACI-Bench-derived response task~\citep{acibench},
HealthDial~\citep{healthdial}, and the public ThReadMed-QA review
sample~\citep{threadmedqa}. The adapted tasks use speech for the latest user turn,
text for history, and hidden references. MedVoice-AID tests responses to medical
audio; MedVoice-Dial tests medical dialogue across specialties. Both benchmarks and
the three adapted tasks use GPT-5.4 signed instance-rubric scoring.

General speech retention is assessed across the backbone, SFT, and full model
on LibriSpeech test-clean~\citep{panayotov2015librispeech} and
AISHELL-1~\citep{bu2017aishell} (WER/CER), VoiceBench-IFEval~\citep{chen2024voicebench}
(official score), and VoiceBench-CommonEval, OpenHermes-Audio
test~\citep{wang2024audiobench}, and a task-stratified WildSpeech-Bench
subset~\citep{zhang2025wildspeechbenchbenchmarkingendtoendspeechllms}
(mean of three independent GPT-5.5 judgments). Inputs, prompts, and greedy
decoding are fixed across stages.
We compare seven general-purpose open-weight baselines and three closed-source
models (Tables~\ref{tab:main_results} and~\ref{tab:open_dialogue_results}), with
matched task content and scoring within each setting. Baselines and input
details appear in Appendix~\ref{app:training_details}.

\subsection{Results}
\label{sec:main_results}

\begin{figure}[!htbp]
\centering
\begin{minipage}[t]{0.45\linewidth}
\vspace{0pt}
\centering
\includegraphics[width=\linewidth]{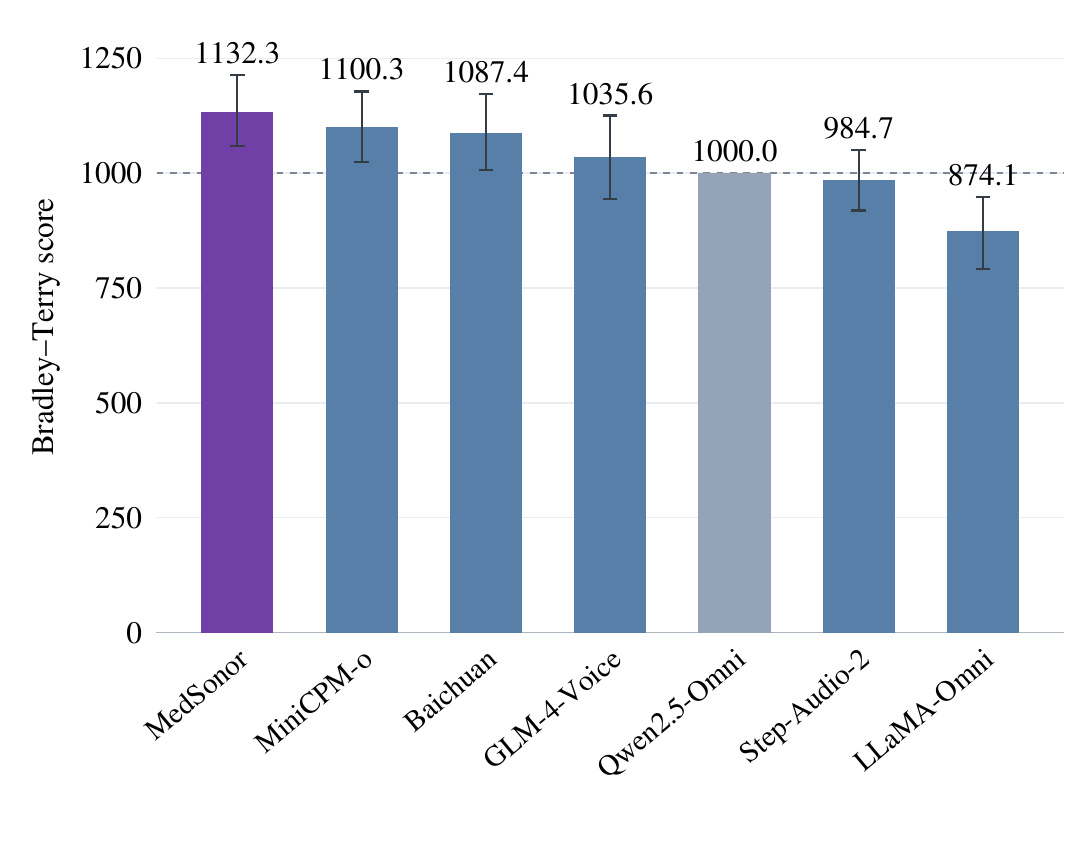}
\end{minipage}\hfill
\begin{minipage}[t]{0.45\linewidth}
\vspace{0pt}
\centering
\includegraphics[width=\linewidth]{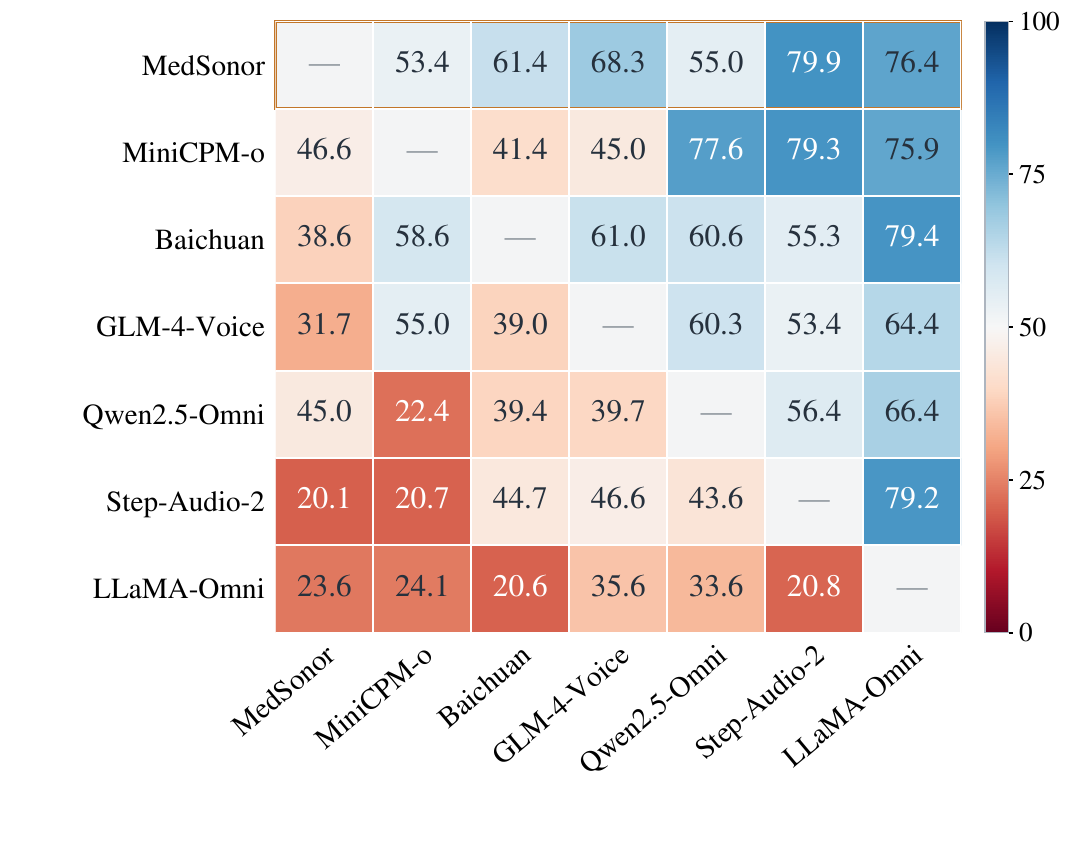}
\end{minipage}
\caption{Physician preferences on 30 MedVoice-AID cases, based on final judgments
from two attending and four resident physicians. Left: aggregate preference
scores with 95\% uncertainty intervals. Right: direct preference credits (\%)
for the row model against the column model, with matching model order.
Colors span 0--100\%, centered at 50\%. Scoring and uncertainty estimation appear in Appendix~\ref{app:human_evaluation}.}
\label{fig:human_eval_arena}
\end{figure}

\noindent  \textbf{Human evaluation.}
\label{sec:human_evaluation}
We conducted a physician preference study on 30 MedVoice-AID cases across five
clinical specialties. Two attending physicians and four physicians in
standardized residency training compared \modelname{} with six open-weight
baselines. Given the audio and two anonymous responses, physicians selected a
preferred response, a tie, or an unjudgeable comparison. Preferences are
aggregated into Bradley--Terry scores with case-bootstrap uncertainty
(Appendix~\ref{app:human_evaluation}). Across 18 comparisons shared by all six physicians, pairwise reviewer
agreement was 59.6\% (95\% item-bootstrap interval: 46.0--73.8\%).
Four comparisons received unanimous judgments from all six physicians. Figure~\ref{fig:human_eval_arena} shows
that \modelname{} ranks above all six baselines, suggesting that our pipeline
improves alignment with the participating physicians' preferences.

\noindent  \textbf{Benchmark performance.}
\modelname{} ranks first among the evaluated open-weight systems on
LiveMedBench Clean-TTS, MMedFD-Lite, MedVoice-AID, and MedVoice-Dial
(Table~\ref{tab:main_results}). These evaluations cover spoken medical
dialogue and responses to clinical requests containing medical audio.
The consistent gains over the backbone suggest that our training pipeline
strengthens medical response generation across both dialogue contexts and
medical audio inputs.

\begin{table}[!htbp]
\centering
\caption{Results on medical dialogue benchmarks.
MMedFD-Lite values are GPT-5.4 G-Eval scores; LiveMedBench Clean-TTS,
MedVoice-AID and MedVoice-Dial values are GPT-5.4
rubric scores. Bold and underline indicate the best and second-best results within the closed-source and open-weight groups separately. The proposed model is included in the open-weight ranking. }
\label{tab:main_results}
\small
\renewcommand{\arraystretch}{0.95}
\setlength{\aboverulesep}{1.5pt}
\setlength{\belowrulesep}{1.5pt}
\setlength{\tabcolsep}{3.5pt}
\begin{lrbox}{\resulttablebox}
\begin{tabular}{l cccc}
\toprule

\textbf{Model}
& \textbf{LiveMedBench Clean-TTS $\uparrow$}
& \textbf{MMedFD-Lite $\uparrow$}
& \textbf{MedVoice-AID $\uparrow$}
& \textbf{MedVoice-Dial $\uparrow$} \\

\midrule

\rowcolor{gray!15}
\multicolumn{5}{c}{\textbf{\textit{Closed-source models}}} \\

Gemini-2.5-Flash
& \underline{0.4079} & 4.624 & 0.1676 & 0.1585 \\

Gemini-3.7-Flash
& \textbf{0.4536} & \textbf{6.872} & \textbf{0.5579} & \textbf{0.7126} \\

GPT-Audio (2025-08-28)
& 0.3798 & \underline{5.224} & \underline{0.5319} & \underline{0.6714} \\

\midrule

\rowcolor{gray!15}
\multicolumn{5}{c}{\textbf{\textit{Open-weight baselines}}} \\

Qwen2.5-Omni-7B
& 0.2593 & 5.532 & 0.3341 & 0.3237 \\

MiniCPM-o-2.6
& 0.2283 & 5.440 & \underline{0.3903} & 0.3307 \\

Baichuan-Omni-1.5
& \underline{0.2803} & \underline{5.692} & 0.3490 & \underline{0.3558} \\

GLM-4-Voice-9B
& 0.2218 & 4.952 & 0.2839 & 0.2902 \\

Step-Audio-2-mini
& 0.2358 & 5.604 & 0.3442 & 0.3367 \\

LLaMA-3.1-8B-Omni
& 0.1122 & 2.556 & 0.1411 & 0.1772 \\

SpeechGPT-2.0-preview
& 0.0688 & 4.272 & -0.1349 & -0.0106 \\

\multicolumn{5}{c}{\textbf{\textit{Proposed model}}} \\
\modelname{}
& \textbf{0.2843}
& \textbf{5.780}
& \textbf{0.7900}
& \textbf{0.5225} \\
\bottomrule
\end{tabular}
\end{lrbox}
\ifdim\wd\resulttablebox>\linewidth
  \resizebox{\linewidth}{!}{\usebox{\resulttablebox}}%
\else
  \usebox{\resulttablebox}%
\fi
\end{table}
On the adapted response evaluations using ACI-Bench, HealthDial, and ThReadMed-QA,
\modelname{} also ranks first among the evaluated open-weight systems
(Table~\ref{tab:open_dialogue_results}). It outperforms Gemini-2.5-Flash on
HealthDial and ThReadMed-QA while narrowing the gap with stronger closed-source
models. These results suggest that the gains extend to medical responses based
on clinical encounters, knowledge-grounded dialogues, and patient--physician
threads, supporting the model's ability to address diverse medical requests
through spoken interaction.
\begin{table}[!htbp]
\centering
\caption{Results on public medical dialogue datasets adapted for spoken response evaluation.
ACI-Bench denotes our dialogue-response task, not its original note-generation benchmark.
All values are GPT-5.4 signed instance-rubric scores. Bold and underline indicate the best and second-best results within the closed-source and open-weight
groups separately. The proposed model is included in the open-weight ranking.}
\label{tab:open_dialogue_results}
\small
\renewcommand{\arraystretch}{0.95}
\setlength{\aboverulesep}{1.5pt}
\setlength{\belowrulesep}{1.5pt}
\setlength{\tabcolsep}{6pt}
\begin{lrbox}{\resulttablebox}
\begin{tabular}{l ccc}
\toprule
 
\textbf{Model}
& \textbf{ACI-Bench $\uparrow$}
& \textbf{HealthDial $\uparrow$}
& \textbf{ThReadMed-QA $\uparrow$} \\

\midrule

\rowcolor{gray!15}
\multicolumn{4}{c}{\textbf{\textit{Closed-source models}}} \\

Gemini-2.5-Flash
& 0.2509 & 0.4661 & \underline{0.2973} \\

Gemini-3.7-Flash
& \textbf{0.4026} & \textbf{0.5875} & \textbf{0.4915} \\

GPT-Audio (2025-08-28)
& \underline{0.3237} & \underline{0.5569} & 0.2858 \\

\midrule

\rowcolor{gray!15}
\multicolumn{4}{c}{\textbf{\textit{Open-weight baselines}}} \\

Qwen2.5-Omni-7B
& 0.2002 & \underline{0.4489} & 0.1800 \\

MiniCPM-o-2.6
& \underline{0.2132} & 0.4160 & 0.1835 \\

Baichuan-Omni-1.5
& 0.1977 & 0.4109 & \underline{0.2100} \\

GLM-4-Voice-9B
& 0.1954 & 0.3678 & 0.1393 \\

Step-Audio-2-mini
& 0.2080 & 0.3605 & 0.1912 \\

LLaMA-3.1-8B-Omni
& 0.0989 & 0.3237 & 0.0845 \\

SpeechGPT-2.0-preview
& -0.0370 & -0.0001 & -0.0084 \\

\multicolumn{4}{c}{\textbf{\textit{Proposed model}}} \\
\modelname{}
& \textbf{0.2263}
& \textbf{0.5022}
& \textbf{0.3403} \\

\bottomrule
\end{tabular}
\end{lrbox}
\ifdim\wd\resulttablebox>\linewidth
  \resizebox{\linewidth}{!}{\usebox{\resulttablebox}}%
\else
  \usebox{\resulttablebox}%
\fi
\end{table}

\begin{figure}[!htbp]
\centering
\begin{minipage}[t]{0.49\linewidth}
\vspace{0pt}
\centering
\includegraphics[width=\linewidth]{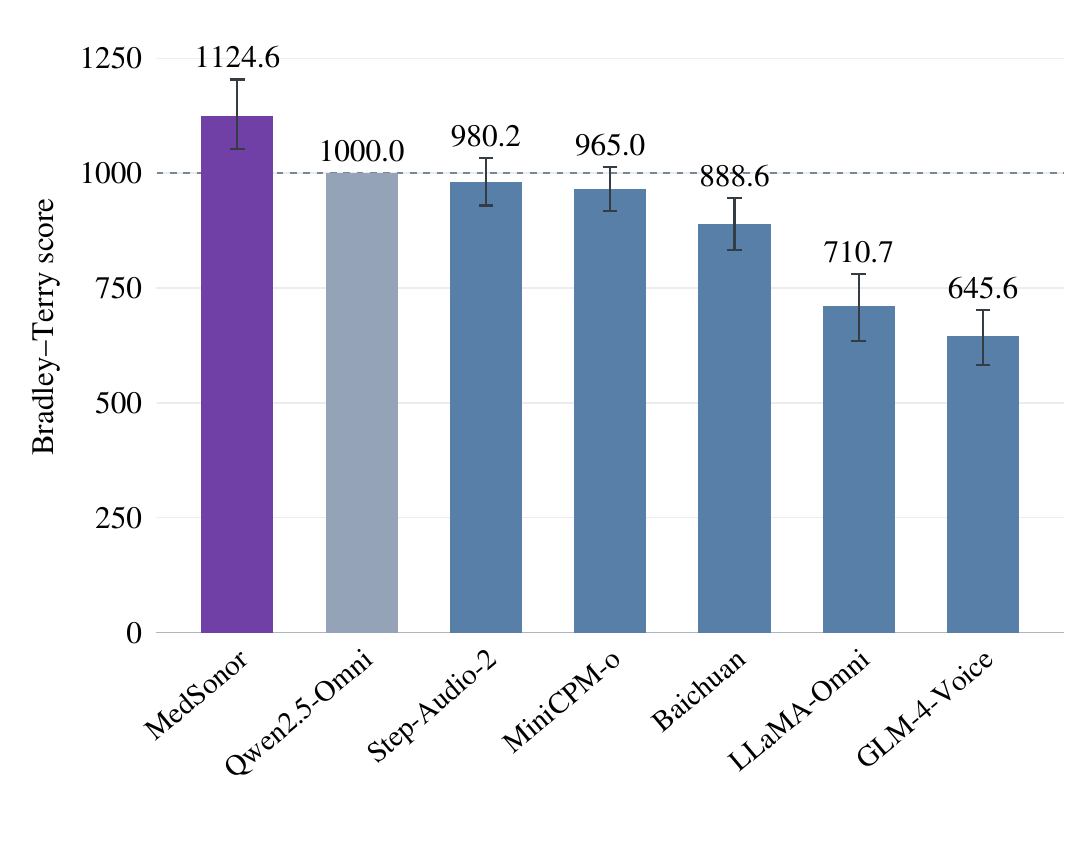}
\end{minipage}\hfill
\begin{minipage}[t]{0.49\linewidth}
\vspace{0pt}
\centering
\includegraphics[width=\linewidth]{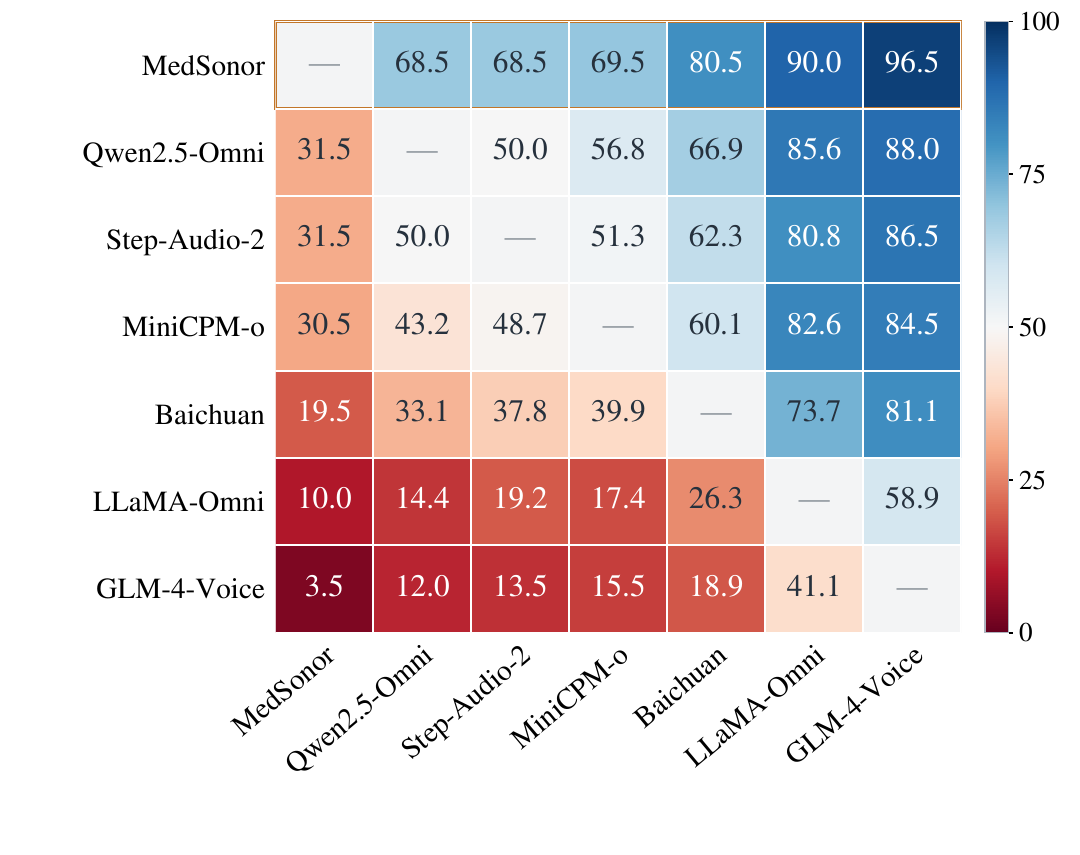}
\end{minipage}
\caption{Arena-style preference evaluation on 100 MedVoice-AID examples from
47 case families, judged by Gemini-3.8-Flash. Left: Bradley--Terry scores
with 95\% intervals from 2,000 case-family bootstrap resamples;
Qwen2.5-Omni-7B is fixed at 1000. Right: direct preference credits (\%) for
the row model against the column model, with model order matched across
both panels. The heatmap colorbar spans 0--100\%, centered at 50\%; colors
do not indicate statistical significance. Ties receive half credit, and
the two presentation orders are averaged.}
\label{fig:medvoice_aid_arena}
\end{figure}

\noindent  \textbf{Pairwise preference evaluation.}
We conducted an automated preference study on 100 MedVoice-AID examples,
using Gemini-3.8-Flash to compare responses from \modelname{} and six
open-weight baselines. The evaluator received the same user audio and two
responses without model identifiers, selecting the preferred response based
on accuracy, grounding, usefulness, and clarity, with ties allowed.
Each pair was evaluated in both presentation orders. We averaged these
judgments and used the same ranking method as in the physician study,
with uncertainty estimated by case-family bootstrap resampling.
As shown in Figure~\ref{fig:medvoice_aid_arena}, \modelname{} ranks above all
six baselines in the aggregate preference ranking. This result complements
the rubric-based evaluation by showing that \modelname{}'s advantage is also
reflected in judgments of complete medical responses under a different evaluator.

\noindent  \textbf{General speech capability retention.}
Table~\ref{tab:general_speech_retention} summarizes performance on general speech tasks after medical post-training.
  \modelname{} broadly retains the backbone's capabilities across speech recognition, spoken dialogue, and open-ended
  instruction following. Performance improves on several dialogue and instruction tasks, while modest declines occur on English
  ASR and OpenHermes-Audio. These results suggest that medical audio dialogue training strengthens domain-specific capabilities
  while largely preserving general speech performance.

\begin{table}[!htbp]
\centering
\setlength{\belowcaptionskip}{4pt}
\caption{General speech capability retention after medical post-training.
Libri and AISHELL report WER/CER (\%;
lower is better); IFEval reports \%; CommonEval and OpenHermes report scores
out of 5; WildSpeech reports a score out of 10. GPT-5.5 scores average three
independent judgments.}
\label{tab:general_speech_retention}
\small
\renewcommand{\arraystretch}{1.05}
\setlength{\tabcolsep}{5pt}
\begin{tabular}{lcccccc}
\toprule
& \multicolumn{2}{c}{\textbf{ASR}}
& \multicolumn{2}{c}{\textbf{Dialogue}}
& \multicolumn{2}{c}{\textbf{Open-ended instruction}} \\
\cmidrule(lr){2-3}\cmidrule(lr){4-5}\cmidrule(l){6-7}
\textbf{Model}
& \textbf{Libri}
& \textbf{AISHELL}
& \textbf{IFEval}
& \textbf{CommonEval}
& \textbf{OpenHermes}
& \textbf{WildSpeech} \\
\midrule
\baseModel{} & 2.26 & 2.93 & 35.55 & 1.77 & 3.49 & 3.35 \\
\modelname{} (SFT) & 2.60 & 2.13 & 47.58 & 3.40 & 3.47 & 4.87 \\
\modelname{} & 2.78 & 1.73 & 45.34 & 3.30 & 3.36 & 4.76 \\
\bottomrule
\end{tabular}
\end{table}

\subsection{Ablation Study}
\label{sec:ablation_study}

\begin{figure}[!htbp]
\centering
\includegraphics[width=\linewidth]{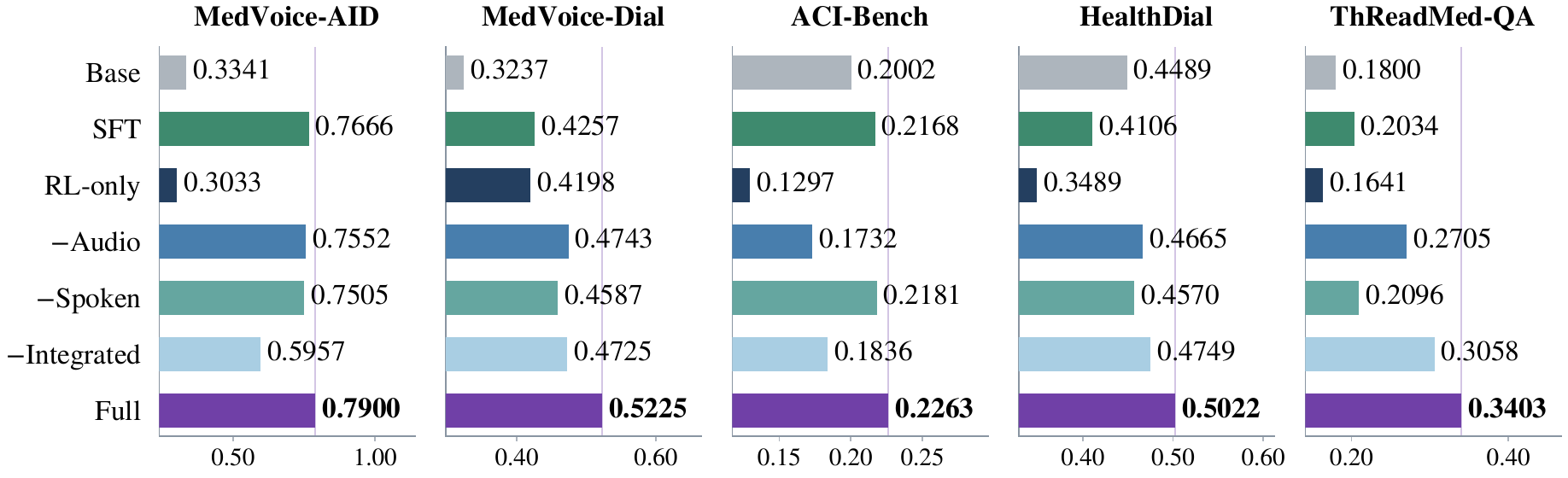}
\caption{Ablations of medical SFT and training sources. Base is Qwen2.5-Omni-7B. SFT and RL-only omit RL and medical SFT, respectively. Source ablations remove one SFT source while retaining both RL stages. Panels report GPT-5.4 rubric scores (higher is better) with independently scaled, truncated axes. Purple lines indicate the full pipeline. Complete numerical results are reported in
Appendix~\ref{app:ablation_results}, Table~\ref{tab:rl_only_ablation}.}
\label{fig:sft_source_ablation}
\end{figure}

\noindent  \textbf{Contribution of multimodal medical SFT.}
  We compare the backbone, SFT-only model, and full pipeline to examine how medical response capabilities develop through
  successive SFT and RL stages. An RL-only variant applies both RL stages directly to Qwen2.5-Omni-7B to assess the
  contribution of SFT initialization. As shown in Figure~\ref{fig:sft_source_ablation} and Table~\ref{tab:rl_only_ablation},
  SFT improves on the backbone on four of the five tasks, and the full pipeline further improves on SFT across all five. This
  pattern supports gains from medical SFT and subsequent RL. RL-only exceeds the backbone only on MedVoice-Dial and falls below the full pipeline on every task, supporting
  the value of medical SFT before reinforcement learning.

\noindent  \textbf{Contributions of spoken dialogue and collected audio.}
  We remove synthesized spoken dialogue and collected medical audio separately to examine their contributions to medical
  response generation. As shown in Figure~\ref{fig:sft_source_ablation} and Table~\ref{tab:rl_only_ablation}, removing spoken
  dialogue most strongly affects tasks requiring responses to conversation context, supporting its role in contextual medical
  interaction. Removing collected medical audio reduces performance across all evaluated tasks, suggesting that this
  supervision benefits medical response generation across different settings. These findings support combining spoken dialogue
  and collected audio in medical SFT.

\noindent  \textbf{Contribution of audio-integrated dialogue.}
  We remove audio-integrated dialogue supervision to examine its contribution to answering clinical requests that involve
  medical audio. As shown in Figure~\ref{fig:sft_source_ablation} and Table~\ref{tab:rl_only_ablation}, its removal
  produces the largest performance decline on MedVoice-AID, supporting the value of explicitly training the model to connect
  audio with clinical requests. Together with the spoken-dialogue ablation, this result suggests that complementary
  supervision formats contribute to the model's ability to handle both medical audio and dialogue context.

\subsection{Robustness to Background Noise}
\label{sec:noise_robustness}

Real-world clinical encounters often involve background noise, such as
nearby conversations, that can interfere with spoken requests and medical
audio. To evaluate robustness to this interference, we overlay babble
noise from the MS-SNSD test set~\citep{reddy2019mssnsd} on the complete input
audio. We select one example from each of 100 independent audio sources in
MedVoice-AID and use identical noise segments across models. After matching
the sampling rate and duration, we mix the input $x$ and zero-mean noise $n$ as
\begin{equation}
    x_{\alpha}=x+\alpha\frac{\operatorname{RMS}(x)}{\operatorname{RMS}(n)}n,
\end{equation}
where $\alpha\in\{0.2,0.6\}$ controls the noise amplitude, corresponding to
approximately 14.0 and 4.4\,dB whole-input signal-to-noise ratios,
respectively. We evaluate medical responses with the same GPT-5.4 signed
instance-rubric metric and greedy decoding.

As shown in Table~\ref{tab:noise_robustness}, MedSonor achieves the highest medical response scores at both noise levels. Although all evaluated models score lower under strong noise than under moderate noise, MedSonor maintains relatively
  stable performance. These results demonstrate its robustness to background noise.

\begin{table}[!htbp]
\centering
\caption{Medical response quality under MS-SNSD babble noise (higher is
better). }
\label{tab:noise_robustness}
\small
\setlength{\tabcolsep}{5pt}
\begin{tabular}{lccc}
\toprule
\textbf{Model} & $\textbf{Clean} (\boldsymbol{\alpha=0})$ & \textbf{Moderate noise}($\boldsymbol{\alpha=0.2})$ & $\textbf{Strong noise} (\boldsymbol{\alpha=0.6})$ \\
\midrule
Qwen2.5-Omni-7B & 0.3341 & 0.2554 & 0.2212 \\
Baichuan-Omni-1.5 & 0.3490 & 0.3039 & 0.2140 \\
GLM-4-Voice-9B & 0.2839 & 0.2310 & 0.2076 \\
Step-Audio-2-mini & 0.3442 & 0.3173 & 0.2645 \\
LLaMA-3.1-8B-Omni & 0.1411 & 0.1858 & 0.1560 \\
\modelname{} & 0.7900 & 0.7480 & 0.7282 \\
\bottomrule
\end{tabular}
\end{table}

\section{Related Work}
\label{sec:related_work}

\noindent \textbf{Medical models.}
Med-PaLM studies medical question answering, while MEDITRON adapts language
models through pretraining on medical literature and guidelines~\citep{medpalm,meditron}.
HuatuoGPT and DISC-MedLLM use medical dialogue supervision~\citep{huatuogpt,discmedllm},
and AMIE develops diagnostic dialogue through simulated
consultations~\citep{amie}. Med-PaLM M extends biomedical modeling to imaging
and genomics~\citep{medpalm_m}. These studies advance medical knowledge and
interaction, but their reported tasks do not directly address clinical
acoustic evidence in spoken consultation.

\noindent \textbf{Medical audio.}
Qwen2-Audio and Qwen2.5-Omni support general audio understanding and
interaction~\citep{qwen2audio,qwen25omni}, while HeAR and OPERA learn
representations for health and respiratory acoustics~\citep{hear2024,opera2024}.
SpeechMedAssist connects spoken consultation with cough and other vocal
cues~\citep{speechmedassist}. StethoLM uses supervised and preference-based
training for cardiopulmonary tasks, including reporting and
reasoning~\citep{stetholm}. We consider diverse clinical sounds and emphasize
how their interpretation should inform responses to different requests
and consultation contexts.

\noindent \textbf{Rubric-based learning.}
HealthBench and MedDialogRubrics provide detailed criteria for evaluating
medical responses and multi-turn consultations~\citep{healthbench,meddialogrubrics}.
Rubrics as Rewards turns instance-specific criteria into feedback for
open-ended reinforcement learning~\citep{rubricsasrewards}, while AudioRubrics
uses evolving, audio-grounded criteria for audio reasoning~\citep{audiorubrics}.
We jointly assess acoustic interpretation and clinical appropriateness
with local item-level judges. Severity-based reward caps prevent other
strengths from offsetting detected critical errors.

\section{Conclusion}
\label{sec:conclusion}
We presented \modelname{}, which combines multimodal medical SFT with two-stage
rubric-guided RL, and introduced MedVoice-Dial and MedVoice-AID for contextual
medical dialogue and clinical audio evaluation.
Across seven medical benchmarks, \modelname{} outperforms the evaluated
general-purpose open-weight audio-language baselines and achieves the highest
aggregate physician preference ranking on 30 MedVoice-AID cases.
Ablations support medical SFT before RL and combined audio--dialogue supervision,
while evaluations show broadly retained speech capabilities and relative
stability under babble noise. These results support adapting medical voice
assistants through supervision that connects audio evidence to clinical requests.

\noindent \textbf{Limitation}
Our work has several limitations. Our datasets are English-only, requiring extension to other languages and dialects and validation across regional clinical settings. Rubric-based evaluation relies on GPT-5.4, and larger-scale physician evaluation remains necessary despite supporting physician preference judgments. The collected-audio SFT stream includes reconstructed Bridge2AI-Voice waveforms, whose acoustic fidelity has not been clinically validated. MedSonor is for research use only and must not be used for clinical diagnosis without rigorous prospective validation and appropriate clinical oversight.

\subsection*{Ethics and Safety Considerations}
\label{sec:ethics}
Source datasets are distributed through public repositories or controlled-access
portals and remain subject to their respective licenses and access conditions.
Use requires qualified healthcare oversight, with professionals retaining responsibility for clinical decisions.
Deployment requires audio quality and completeness checks, systematic output review for medical accuracy,
unsupported claims, and harmful advice, plus ongoing monitoring and escalation of uncertain or high-risk cases to professionals.

\subsection*{AI Use Statement}
We used LLMs to assist with language editing, data synthesis and code implementation. 
LLM-generated code was verified and tested for correctness.
We take responsibility for the final content of this work,
including text, claims or artifacts produced with the aid of generative AI.

{\small
\bibliographystyle{unsrtnat}

\bibliography{ref}

\begin{thebibliography}{57}
\providecommand{\natexlab}[1]{#1}
\providecommand{\url}[1]{\texttt{#1}}
\expandafter\ifx\csname urlstyle\endcsname\relax
  \providecommand{\doi}[1]{doi: #1}\else
  \providecommand{\doi}{doi: \begingroup \urlstyle{rm}\Url}\fi

\bibitem[Rocha et~al.(2019)Rocha, Filos, Mendes, Serbes, Ulukaya, Kahya,
  Jakovljevic, Turukalo, Vogiatzis, Perantoni, Kaimakamis, Natsiavas, Oliveira,
  J{\'a}come, Marques, Maglaveras, Paiva, Chouvarda, and
  de~Carvalho]{rocha2019respiratory}
Bruno~M. Rocha, Dimitris Filos, Lu{\'i}s Mendes, Gorkem Serbes, Sezer Ulukaya,
  Yasemin~P. Kahya, Nik{\v{s}}a Jakovljevic, Tatjana~L. Turukalo, Ioannis~M.
  Vogiatzis, Eleni Perantoni, Evangelos Kaimakamis, Pantelis Natsiavas, Ana
  Oliveira, Cristina J{\'a}come, Alda Marques, Nicos Maglaveras, Rui~Pedro
  Paiva, Ioanna Chouvarda, and Paulo de~Carvalho.
\newblock An open access database for the evaluation of respiratory sound
  classification algorithms.
\newblock \emph{Physiological Measurement}, 40\penalty0 (3):\penalty0 035001,
  2019.
\newblock \doi{10.1088/1361-6579/ab03ea}.
\newblock URL \url{https://doi.org/10.1088/1361-6579/ab03ea}.

\bibitem[Oliveira et~al.(2022)Oliveira, Renna, Costa, Nogueira, Oliveira,
  Ferreira, Jorge, Mattos, Hatem, Tavares, Elola, {Bahrami Rad}, Sameni,
  Clifford, and Coimbra]{oliveira2022circor}
Jorge Oliveira, Francesco Renna, Paulo~Dias Costa, Marcelo Nogueira, Cristina
  Oliveira, Carlos Ferreira, Alipio Jorge, Sandra Mattos, Thamine Hatem, Thiago
  Tavares, Andoni Elola, Ali {Bahrami Rad}, Reza Sameni, Gari~D. Clifford, and
  Miguel~T. Coimbra.
\newblock The {CirCor DigiScope} dataset: From murmur detection to murmur
  classification.
\newblock \emph{IEEE Journal of Biomedical and Health Informatics}, 26\penalty0
  (6):\penalty0 2524--2535, 2022.
\newblock \doi{10.1109/JBHI.2021.3137048}.
\newblock URL \url{https://doi.org/10.1109/JBHI.2021.3137048}.

\bibitem[Bensoussan et~al.(2024)Bensoussan, Ghosh, Rameau, Boyer, Bahr, Watts,
  Rudzicz, Bolser, Lerner-Ellis, Awan, Powell, Belisle-Pipon, Ravitsky,
  Johnson, Sigaras, Elemento, Dorr, and Payne]{bensoussan2024voice}
Yael Bensoussan, Satrajit Ghosh, Anais Rameau, Micah Boyer, Ruth Bahr,
  Stephanie Watts, Frank Rudzicz, Don Bolser, Jordan Lerner-Ellis, Shaheen
  Awan, Maria Powell, Jean-Christophe Belisle-Pipon, Vardit Ravitsky, Alistair
  Johnson, Alexandros Sigaras, Olivier Elemento, David Dorr, and Philip Payne.
\newblock Developing multi-disorder voice protocols: A team science approach
  involving clinical expertise, bioethics, standards, and {DEI}.
\newblock In \emph{Interspeech 2024}, pages 1445--1449, 2024.
\newblock \doi{10.21437/Interspeech.2024-1926}.
\newblock URL
  \url{https://www.isca-archive.org/interspeech_2024/bensoussan24_interspeech.html}.

\bibitem[Singhal et~al.(2023)Singhal, Azizi, Tu, Mahdavi, Wei, Chung, Scales,
  Tanwani, Cole-Lewis, Pfohl, et~al.]{medpalm}
Karan Singhal, Shekoofeh Azizi, Tao Tu, S.~Sara Mahdavi, Jason Wei, Hyung~Won
  Chung, Nathan Scales, Ajay Tanwani, Heather Cole-Lewis, Stephen Pfohl, et~al.
\newblock Large language models encode clinical knowledge.
\newblock \emph{Nature}, 620:\penalty0 172--180, 2023.
\newblock \doi{10.1038/s41586-023-06291-2}.
\newblock URL \url{https://www.nature.com/articles/s41586-023-06291-2}.

\bibitem[Chen et~al.(2023)Chen, Hern{\'a}ndez~Cano, Romanou, Bonnet, Matoba,
  Salvi, Pagliardini, Fan, K{\"o}pf, Mohtashami, Sallinen, Sakhaeirad, Swamy,
  Krawczuk, Bayazit, Marmet, Montariol, Hartley, Jaggi, and Bosselut]{meditron}
Zeming Chen, Alejandro Hern{\'a}ndez~Cano, Angelika Romanou, Antoine Bonnet,
  Kyle Matoba, Francesco Salvi, Matteo Pagliardini, Simin Fan, Andreas
  K{\"o}pf, Amirkeivan Mohtashami, Alexandre Sallinen, Alireza Sakhaeirad,
  Vinitra Swamy, Igor Krawczuk, Deniz Bayazit, Axel Marmet, Syrielle Montariol,
  Mary-Anne Hartley, Martin Jaggi, and Antoine Bosselut.
\newblock {MEDITRON-70B}: Scaling medical pretraining for large language
  models.
\newblock \emph{arXiv preprint arXiv:2311.16079}, 2023.
\newblock URL \url{https://arxiv.org/abs/2311.16079}.

\bibitem[Tu et~al.(2024)Tu, Palepu, Schaekermann, Saab, Freyberg, Tanno, Wang,
  Li, Amin, Tomasev, et~al.]{amie}
Tao Tu, Anil Palepu, Mike Schaekermann, Khaled Saab, Jan Freyberg, Ryutaro
  Tanno, Amy Wang, Brenna Li, Mohamed Amin, Nenad Tomasev, et~al.
\newblock Towards conversational diagnostic {AI}.
\newblock \emph{arXiv preprint arXiv:2401.05654}, 2024.
\newblock URL \url{https://arxiv.org/abs/2401.05654}.

\bibitem[Dao et~al.(2025)Dao, Schneiders, Williams, Bautista, Seabrooke,
  Vigneswaran, Kolpekwar, Vashistha, and Farahi]{dao2025tame}
Tu-Quyen Dao, Eike Schneiders, Jennifer Williams, John~Robert Bautista, Tina
  Seabrooke, Ganesh Vigneswaran, Rishik Kolpekwar, Ritwik Vashistha, and Arya
  Farahi.
\newblock {TAME Pain}: Trustworthy {AssessMEnt} of pain from speech and audio
  for the empowerment of patients.
\newblock \emph{PhysioNet}, 2025.
\newblock \doi{10.13026/20e2-1g10}.
\newblock URL \url{https://physionet.org/content/tame-pain/1.0.0/}.
\newblock Version 1.0.0.

\bibitem[Bellieni et~al.(2004)Bellieni, Sisto, Cordelli, and
  Buonocore]{bellieni2004cry}
Carlo~V. Bellieni, Renata Sisto, Duccio~M. Cordelli, and Giuseppe Buonocore.
\newblock Cry features reflect pain intensity in term newborns: An alarm
  threshold.
\newblock \emph{Pediatric Research}, 55:\penalty0 142--146, 2004.
\newblock \doi{10.1203/01.PDR.0000099793.99608.CB}.
\newblock URL \url{https://www.nature.com/articles/pr200421}.

\bibitem[Baur et~al.(2024)Baur, Nabulsi, Weng, Garrison, Blankemeier, Fishman,
  Chen, Kakarmath, Maimbolwa, Sanjase, Shuma, Matias, Corrado, Patel, Shetty,
  Prabhakara, Muyoyeta, and Ardila]{hear2024}
Sebastien Baur, Zaid Nabulsi, Wei-Hung Weng, Jake Garrison, Louis Blankemeier,
  Sam Fishman, Christina Chen, Sujay Kakarmath, Minyoi Maimbolwa, Nsala
  Sanjase, Brian Shuma, Yossi Matias, Greg~S. Corrado, Shwetak Patel, Shravya
  Shetty, Shruthi Prabhakara, Monde Muyoyeta, and Diego Ardila.
\newblock {HeAR} -- health acoustic representations.
\newblock \emph{arXiv preprint arXiv:2403.02522}, 2024.
\newblock URL \url{https://arxiv.org/abs/2403.02522}.

\bibitem[Xu et~al.(2025)Xu, Guo, He, Hu, He, Bai, Chen, Wang, Fan, Dang, Zhang,
  Wang, Chu, and Lin]{qwen25omni}
Jin Xu, Zhifang Guo, Jinzheng He, Hangrui Hu, Ting He, Shuai Bai, Keqin Chen,
  Jialin Wang, Yang Fan, Kai Dang, Bin Zhang, Xiong Wang, Yunfei Chu, and
  Junyang Lin.
\newblock Qwen2.5-omni technical report.
\newblock \emph{arXiv preprint arXiv:2503.20215}, 2025.
\newblock URL \url{https://arxiv.org/abs/2503.20215}.

\bibitem[Chen et~al.(2026)Chen, Wang, Chen, and Wei]{speechmedassist}
Sirry Chen, Jieyi Wang, Wei Chen, and Zhongyu Wei.
\newblock {S}peech{M}ed{A}ssist: Efficiently and effectively adapting speech
  language models for medical consultation.
\newblock In \emph{Proceedings of the 64th Annual Meeting of the Association
  for Computational Linguistics (Volume 1: Long Papers)}, pages 30914--30935,
  2026.
\newblock \doi{10.18653/v1/2026.acl-long.1428}.
\newblock URL \url{https://aclanthology.org/2026.acl-long.1428/}.

\bibitem[Wang et~al.(2026)Wang, Wang, Funk, and Saeed]{stetholm}
Yishan Wang, Tsai-Ning Wang, Mathias Funk, and Aaqib Saeed.
\newblock {StethoLM}: Audio language model for cardiopulmonary analysis across
  clinical tasks.
\newblock \emph{arXiv preprint arXiv:2603.00355}, 2026.
\newblock URL \url{https://arxiv.org/abs/2603.00355}.

\bibitem[Yu et~al.(2025)Yu, Zhang, Zhu, Yuan, Zuo, Yue, Dai, Fan, Liu, Liu,
  Liu, Lin, Lin, Ma, Sheng, Tong, Zhang, Zhang, Zhang, Zhu, Zhu, Chen, Chen,
  Wang, Yu, Song, Wei, Zhou, Liu, Ma, Zhang, Yan, Qiao, Wu, and Wang]{dapo}
Qiying Yu, Zheng Zhang, Ruofei Zhu, Yufeng Yuan, Xiaochen Zuo, Yu~Yue, Weinan
  Dai, Tiantian Fan, Gaohong Liu, Lingjun Liu, Xin Liu, Haibin Lin, Zhiqi Lin,
  Bole Ma, Guangming Sheng, Yuxuan Tong, Chi Zhang, Mofan Zhang, Wang Zhang,
  Hang Zhu, Jinhua Zhu, Jiaze Chen, Jiangjie Chen, Chengyi Wang, Hongli Yu,
  Yuxuan Song, Xiangpeng Wei, Hao Zhou, Jingjing Liu, Wei-Ying Ma, Ya-Qin
  Zhang, Lin Yan, Mu~Qiao, Yonghui Wu, and Mingxuan Wang.
\newblock Dapo: An open-source llm reinforcement learning system at scale.
\newblock \emph{arXiv preprint arXiv:2503.14476}, 2025.
\newblock URL \url{https://arxiv.org/abs/2503.14476}.

\bibitem[Yan et~al.(2026)Yan, Song, Fang, Ji, Li, Li, and Sun]{livemedbench}
Zhiling Yan, Dingjie Song, Zhe Fang, Yisheng Ji, Xiang Li, Quanzheng Li, and
  Lichao Sun.
\newblock Livemedbench: A contamination-free medical benchmark for llms with
  automated rubric evaluation.
\newblock \emph{arXiv preprint arXiv:2602.10367}, 2026.
\newblock URL \url{https://arxiv.org/abs/2602.10367}.

\bibitem[Chen et~al.(2025{\natexlab{a}})Chen, Wang, Lan, Ding, Jiang, Yang, Xu,
  Luo, Ng, Cheng, Mao, and Yoo]{mmedfd}
Hongzhao Chen, Xiaoyang Wang, Jing Lan, Hexiao Ding, Yufeng Jiang, Minghui
  Yang, Danhui Xu, Jun Luo, Nga-Chun Ng, Gerald W.~Y. Cheng, Yunlin Mao, and
  Jung~Sun Yoo.
\newblock Mmedfd: A real-world healthcare benchmark for multi-turn full-duplex
  automatic speech recognition.
\newblock \emph{arXiv preprint arXiv:2509.19817}, 2025{\natexlab{a}}.
\newblock URL \url{https://arxiv.org/abs/2509.19817}.

\bibitem[Yim et~al.(2023)Yim, Fu, Ben~Abacha, Snider, Lin, and
  Yetisgen]{acibench}
Wen-wai Yim, Yujuan Fu, Asma Ben~Abacha, Neal Snider, Thomas Lin, and Meliha
  Yetisgen.
\newblock {ACI-BENCH}: A novel ambient clinical intelligence dataset for
  benchmarking automatic visit note generation.
\newblock \emph{Scientific Data}, 10\penalty0 (1):\penalty0 586, 2023.
\newblock \doi{10.1038/s41597-023-02487-3}.
\newblock URL \url{https://doi.org/10.1038/s41597-023-02487-3}.

\bibitem[Hu et~al.(2026)Hu, Liu, Zhou, Razumovskaia, Wang, Fraser, Vuli{\'c},
  and Korhonen]{healthdial}
Songbo Hu, Yinhong Liu, Ej~Zhou, Evgeniia Razumovskaia, Xiaobin Wang, Alexander
  Fraser, Ivan Vuli{\'c}, and Anna Korhonen.
\newblock Dial {HEALTHDIAL} for advice: A multilingual and multi-parallel
  spoken dialogue dataset for knowledge-grounded information seeking.
\newblock In \emph{Findings of the Association for Computational Linguistics:
  ACL 2026}, pages 25529--25552, 2026.
\newblock \doi{10.18653/v1/2026.findings-acl.1275}.
\newblock URL \url{https://aclanthology.org/2026.findings-acl.1275/}.

\bibitem[Munnangi and Savage(2026)]{threadmedqa}
Monica Munnangi and Saiph Savage.
\newblock {ThReadMed-QA}: A multi-turn medical dialogue benchmark from real
  patient questions.
\newblock \emph{arXiv preprint arXiv:2603.11281}, 2026.
\newblock URL \url{https://arxiv.org/abs/2603.11281}.

\bibitem[Panayotov et~al.(2015)Panayotov, Chen, Povey, and
  Khudanpur]{panayotov2015librispeech}
Vassil Panayotov, Guoguo Chen, Daniel Povey, and Sanjeev Khudanpur.
\newblock {LibriSpeech}: An {ASR} corpus based on public domain audio books.
\newblock In \emph{2015 IEEE International Conference on Acoustics, Speech and
  Signal Processing (ICASSP)}, pages 5206--5210, 2015.
\newblock \doi{10.1109/ICASSP.2015.7178964}.
\newblock URL \url{https://doi.org/10.1109/ICASSP.2015.7178964}.

\bibitem[Bu et~al.(2017)Bu, Du, Na, Wu, and Zheng]{bu2017aishell}
Hui Bu, Jiayu Du, Xingyu Na, Bengu Wu, and Hao Zheng.
\newblock {AISHELL-1}: An open-source mandarin speech corpus and a speech
  recognition baseline.
\newblock In \emph{2017 20th Conference of the Oriental Chapter of the
  International Coordinating Committee on Speech Databases and Speech I/O
  Systems and Assessment (O-COCOSDA)}, pages 1--5, 2017.
\newblock \doi{10.1109/ICSDA.2017.8384449}.
\newblock URL \url{https://doi.org/10.1109/ICSDA.2017.8384449}.

\bibitem[Chen et~al.(2024)Chen, Yue, Zhang, Gao, Tan, and
  Li]{chen2024voicebench}
Yiming Chen, Xianghu Yue, Chen Zhang, Xiaoxue Gao, Robby~T. Tan, and Haizhou
  Li.
\newblock {VoiceBench}: Benchmarking {LLM}-based voice assistants.
\newblock \emph{arXiv preprint arXiv:2410.17196}, 2024.
\newblock URL \url{https://arxiv.org/abs/2410.17196}.

\bibitem[Wang et~al.(2025)Wang, Zou, Lin, Sun, Liu, Zhang, Liu, Aw, and
  Chen]{wang2024audiobench}
Bin Wang, Xunlong Zou, Geyu Lin, Shuo Sun, Zhuohan Liu, Wenyu Zhang, Zhengyuan
  Liu, AiTi Aw, and Nancy~F. Chen.
\newblock {AudioBench}: A universal benchmark for audio large language models.
\newblock In \emph{Proceedings of the 2025 Conference of the Nations of the
  Americas Chapter of the Association for Computational Linguistics: Human
  Language Technologies (Volume 1: Long Papers)}, pages 4297--4316, 2025.
\newblock \doi{10.18653/v1/2025.naacl-long.218}.
\newblock URL \url{https://aclanthology.org/2025.naacl-long.218/}.

\bibitem[Zhang et~al.(2025)Zhang, Zhang, Lei, Wu, Liu, Jia, and
  Zhou]{zhang2025wildspeechbenchbenchmarkingendtoendspeechllms}
Linhao Zhang, Jian Zhang, Bokai Lei, Chuhan Wu, Aiwei Liu, Wei Jia, and Xiao
  Zhou.
\newblock {WildSpeech-Bench}: Benchmarking end-to-end {SpeechLLMs} in the wild,
  2025.
\newblock URL \url{https://arxiv.org/abs/2506.21875}.

\bibitem[Reddy et~al.(2019)Reddy, Beyrami, Pool, Cutler, Srinivasan, and
  Gehrke]{reddy2019mssnsd}
Chandan K.~A. Reddy, Ebrahim Beyrami, Jamie Pool, Ross Cutler, Sriram
  Srinivasan, and Johannes Gehrke.
\newblock A scalable noisy speech dataset and online subjective test framework.
\newblock \emph{arXiv preprint arXiv:1909.08050}, 2019.
\newblock URL \url{https://arxiv.org/abs/1909.08050}.

\bibitem[Zhang et~al.(2023)Zhang, Chen, Jiang, Yu, Chen, Li, Chen, Wu, Zhang,
  Xiao, Wan, Wang, and Li]{huatuogpt}
Hongbo Zhang, Junying Chen, Feng Jiang, Fei Yu, Zhihong Chen, Jianquan Li,
  Guiming Chen, Xiangbo Wu, Zhiyi Zhang, Qingying Xiao, Xiang Wan, Benyou Wang,
  and Haizhou Li.
\newblock {HuatuoGPT}, towards taming language model to be a doctor.
\newblock \emph{arXiv preprint arXiv:2305.15075}, 2023.
\newblock URL \url{https://arxiv.org/abs/2305.15075}.

\bibitem[Bao et~al.(2023)Bao, Chen, Xiao, Ren, Wu, Zhong, Peng, Huang, and
  Wei]{discmedllm}
Zhijie Bao, Wei Chen, Shengze Xiao, Kuang Ren, Jiaao Wu, Cheng Zhong, Jiajie
  Peng, Xuanjing Huang, and Zhongyu Wei.
\newblock {DISC-MedLLM}: Bridging general large language models and real-world
  medical consultation.
\newblock \emph{arXiv preprint arXiv:2308.14346}, 2023.
\newblock URL \url{https://arxiv.org/abs/2308.14346}.

\bibitem[Tu et~al.(2023)Tu, Azizi, Driess, Schaekermann, Amin, Chang, Carroll,
  Lau, Tanno, Ktena, et~al.]{medpalm_m}
Tao Tu, Shekoofeh Azizi, Danny Driess, Mike Schaekermann, Mohamed Amin,
  Pi-Chuan Chang, Andrew Carroll, Chuck Lau, Ryutaro Tanno, Ira Ktena, et~al.
\newblock Towards generalist biomedical {AI}.
\newblock \emph{arXiv preprint arXiv:2307.14334}, 2023.
\newblock URL \url{https://arxiv.org/abs/2307.14334}.

\bibitem[Chu et~al.(2024)Chu, Xu, Yang, Wei, Wei, Guo, Leng, Lv, He, Lin, Zhou,
  and Zhou]{qwen2audio}
Yunfei Chu, Jin Xu, Qian Yang, Haojie Wei, Xipin Wei, Zhifang Guo, Yichong
  Leng, Yuanjun Lv, Jinzheng He, Junyang Lin, Chang Zhou, and Jingren Zhou.
\newblock {Qwen2-Audio} technical report.
\newblock \emph{arXiv preprint arXiv:2407.10759}, 2024.
\newblock URL \url{https://arxiv.org/abs/2407.10759}.

\bibitem[Zhang et~al.(2024)Zhang, Xia, Han, Wu, Rizos, Liu, Mosuily, Chauhan,
  and Mascolo]{opera2024}
Yuwei Zhang, Tong Xia, Jing Han, Yu~Wu, Georgios Rizos, Yang Liu, Mohammed
  Mosuily, Jagmohan Chauhan, and Cecilia Mascolo.
\newblock Towards open respiratory acoustic foundation models: Pretraining and
  benchmarking.
\newblock \emph{arXiv preprint arXiv:2406.16148}, 2024.
\newblock URL \url{https://arxiv.org/abs/2406.16148}.

\bibitem[Arora et~al.(2025)Arora, Wei, Hicks, Bowman, Qui\~nonero Candela,
  Tsimpourlas, Sharman, Shah, Vallone, Beutel, Heidecke, and
  Singhal]{healthbench}
Rahul~K. Arora, Jason Wei, Rebecca~Soskin Hicks, Preston Bowman, Joaquin
  Qui\~nonero Candela, Foivos Tsimpourlas, Michael Sharman, Meghan Shah, Andrea
  Vallone, Alex Beutel, Johannes Heidecke, and Karan Singhal.
\newblock {HealthBench}: Evaluating large language models towards improved
  human health.
\newblock \emph{arXiv preprint arXiv:2505.08775}, 2025.
\newblock URL \url{https://arxiv.org/abs/2505.08775}.

\bibitem[Gong et~al.(2026)Gong, Fang, Yang, Tao, Guo, Wei, Xie, Guan, Chen,
  Shi, Gu, and Liu]{meddialogrubrics}
Lecheng Gong, Weimin Fang, Ting Yang, Dongjie Tao, Chunxiao Guo, Peng Wei,
  Bo~Xie, Jinqun Guan, Zixiao Chen, Fang Shi, Jinjie Gu, and Junwei Liu.
\newblock {MedDialogRubrics}: A comprehensive benchmark and evaluation
  framework for multi-turn medical consultations in large language models.
\newblock \emph{arXiv preprint arXiv:2601.03023}, 2026.
\newblock URL \url{https://arxiv.org/abs/2601.03023}.

\bibitem[Gunjal et~al.(2025)Gunjal, Wang, Lau, Nath, He, Liu, and
  Hendryx]{rubricsasrewards}
Anisha Gunjal, Anthony Wang, Elaine Lau, Vaskar Nath, Yunzhong He, Bing Liu,
  and Sean Hendryx.
\newblock Rubrics as rewards: Reinforcement learning beyond verifiable domains.
\newblock \emph{arXiv preprint arXiv:2507.17746}, 2025.
\newblock URL \url{https://arxiv.org/abs/2507.17746}.

\bibitem[Yu et~al.(2026)Yu, Feng, Min, Lin, Xu, Xu, Yu, Liu, and
  Zhou]{audiorubrics}
Fangxu Yu, Tao Feng, Dehai Min, Zinan Lin, Weijia Xu, Michael Xu, Philip~S. Yu,
  Ge~Liu, and Tianyi Zhou.
\newblock Reinforcement learning with evolving rubrics as rewards for audio
  reasoning.
\newblock \emph{arXiv preprint arXiv:2608.02831}, 2026.
\newblock URL \url{https://arxiv.org/abs/2608.02831}.

\bibitem[Yao et~al.(2024)Yao, Yu, Zhang, Wang, Cui, Zhu, Cai, Li, Zhao, He,
  et~al.]{minicpmv}
Yuan Yao, Tianyu Yu, Ao~Zhang, Chongyi Wang, Junbo Cui, Hongji Zhu, Tianchi
  Cai, Haoyu Li, Weilin Zhao, Zhihui He, et~al.
\newblock Minicpm-v: A gpt-4v level mllm on your phone.
\newblock \emph{arXiv preprint arXiv:2408.01800}, 2024.
\newblock URL \url{https://arxiv.org/abs/2408.01800}.

\bibitem[Li et~al.(2025)Li, Liu, Zhang, Zhang, Chen, Li, Li, Liu, Ming, Dong,
  et~al.]{baichuanomni}
Yadong Li, Jun Liu, Tao Zhang, Tao Zhang, Song Chen, Tianpeng Li, Zehuan Li,
  Lijun Liu, Lingfeng Ming, Guosheng Dong, et~al.
\newblock Baichuan-omni-1.5 technical report.
\newblock \emph{arXiv preprint arXiv:2501.15368}, 2025.
\newblock URL \url{https://arxiv.org/abs/2501.15368}.

\bibitem[Zeng et~al.(2024)Zeng, Du, Liu, Wang, Jiang, Zhao, Dong, and
  Tang]{glm4voice}
Aohan Zeng, Zhengxiao Du, Mingdao Liu, Kedong Wang, Shengmin Jiang, Lei Zhao,
  Yuxiao Dong, and Jie Tang.
\newblock Glm-4-voice: Towards intelligent and human-like end-to-end spoken
  chatbot.
\newblock \emph{arXiv preprint arXiv:2412.02612}, 2024.
\newblock URL \url{https://arxiv.org/abs/2412.02612}.

\bibitem[Wu et~al.(2025)Wu, Yan, Hu, Yi, Feng, Tian, Shen, Yu, Zhang, Li,
  et~al.]{stepaudio2}
Boyong Wu, Chao Yan, Chen Hu, Cheng Yi, Chengli Feng, Fei Tian, Feiyu Shen,
  Gang Yu, Haoyang Zhang, Jingbei Li, et~al.
\newblock Step-audio 2 technical report.
\newblock \emph{arXiv preprint arXiv:2507.16632}, 2025.
\newblock URL \url{https://arxiv.org/abs/2507.16632}.

\bibitem[Fang et~al.(2024)Fang, Guo, Zhou, Ma, Zhang, and Feng]{llamaomni}
Qingkai Fang, Shoutao Guo, Yan Zhou, Zhengrui Ma, Shaolei Zhang, and Yang Feng.
\newblock Llama-omni: Seamless speech interaction with large language models.
\newblock \emph{arXiv preprint arXiv:2409.06666}, 2024.
\newblock URL \url{https://arxiv.org/abs/2409.06666}.

\bibitem[{Open-Moss}(2025)]{speechgpt2}
{Open-Moss}.
\newblock Speechgpt 2.0-preview.
\newblock GitHub repository, 2025.
\newblock URL \url{https://github.com/OpenMOSS/SpeechGPT-2.0-preview}.

\bibitem[Hsu et~al.(2021)Hsu, Huang, Huang, Huang, Cheng, Chen, Hsiao, Chen,
  Chen, Lai, Hsu, Lin, Tsai, Wu, Tseng, Tseng, Chen, and Lai]{hf_lungsource}
Fu-Shun Hsu, Shang-Ran Huang, Chien-Wen Huang, Chao-Jung Huang, Yuan-Ren Cheng,
  Chun-Chieh Chen, Jack Hsiao, Chung-Wei Chen, Li-Chin Chen, Yen-Chun Lai,
  Bi-Fang Hsu, Nian-Jhen Lin, Wan-Ling Tsai, Yi-Lin Wu, Tzu-Ling Tseng,
  Ching-Ting Tseng, Yi-Tsun Chen, and Feipei Lai.
\newblock {Benchmarking of eight recurrent neural network variants for breath
  phase and adventitious sound detection on a self-developed open-access lung
  sound database---HF\_Lung\_V1}.
\newblock \emph{PLOS ONE}, 16\penalty0 (7):\penalty0 e0254134, 2021.
\newblock \doi{10.1371/journal.pone.0254134}.
\newblock URL \url{https://doi.org/10.1371/journal.pone.0254134}.

\bibitem[Zhang et~al.(2022)Zhang, Zhang, Yuan, Huang, Zhang, Zhang, Lv, Lin,
  Wang, Liu, Tang, Wang, Ma, Liu, Yuan, Zhou, Zhao, Li, Yin, Zhao, Wang, and
  Lian]{sprsoundsource}
Qing Zhang, Jing Zhang, Jiajun Yuan, Huajie Huang, Yuhang Zhang, Baoqin Zhang,
  Gaomei Lv, Shuzhu Lin, Na~Wang, Xin Liu, Mingyu Tang, Yahua Wang, Hui Ma,
  Lu~Liu, Shuhua Yuan, Hongyuan Zhou, Jian Zhao, Yongfu Li, Yong Yin, Liebin
  Zhao, Guoxing Wang, and Yong Lian.
\newblock {SPRSound: Open-Source SJTU Paediatric Respiratory Sound Database}.
\newblock \emph{IEEE Transactions on Biomedical Circuits and Systems},
  16\penalty0 (5):\penalty0 867--881, 2022.
\newblock \doi{10.1109/tbcas.2022.3204910}.
\newblock URL \url{https://doi.org/10.1109/tbcas.2022.3204910}.

\bibitem[Sharma et~al.(2020)Sharma, Krishnan, Kumar, Ramoji, Chetupalli, R.,
  Ghosh, and Ganapathy]{coswarasource}
Neeraj Sharma, Prashant Krishnan, Rohit Kumar, Shreyas Ramoji, Srikanth~Raj
  Chetupalli, Nirmala R., Prasanta~Kumar Ghosh, and Sriram Ganapathy.
\newblock {Coswara --- A Database of Breathing, Cough, and Voice Sounds for
  COVID-19 Diagnosis}.
\newblock In \emph{Interspeech 2020}, pages 4811--4815, 2020.
\newblock \doi{10.21437/interspeech.2020-2768}.
\newblock URL \url{https://doi.org/10.21437/interspeech.2020-2768}.

\bibitem[Orlandic et~al.(2021)Orlandic, Teijeiro, and Atienza]{coughvidsource}
Lara Orlandic, Tomas Teijeiro, and David Atienza.
\newblock {The COUGHVID crowdsourcing dataset, a corpus for the study of
  large-scale cough analysis algorithms}.
\newblock \emph{Scientific Data}, 8\penalty0 (1):\penalty0 156, 2021.
\newblock \doi{10.1038/s41597-021-00937-4}.
\newblock URL \url{https://doi.org/10.1038/s41597-021-00937-4}.

\bibitem[Liu et~al.(2016)Liu, Springer, Li, Moody, et~al.]{physionet2016source}
Chengyu Liu, David Springer, Qiao Li, Benjamin Moody, et~al.
\newblock {An open access database for the evaluation of heart sound
  algorithms}.
\newblock \emph{Physiological Measurement}, 37\penalty0 (12):\penalty0
  2181--2213, 2016.
\newblock \doi{10.1088/0967-3334/37/12/2181}.
\newblock URL \url{https://doi.org/10.1088/0967-3334/37/12/2181}.

\bibitem[Rudzicz et~al.(2012)Rudzicz, Namasivayam, and Wolff]{torgosource}
Frank Rudzicz, Aravind~Kumar Namasivayam, and Talya Wolff.
\newblock {The TORGO database of acoustic and articulatory speech from speakers
  with dysarthria}.
\newblock \emph{Language Resources and Evaluation}, 46\penalty0 (4):\penalty0
  523--541, 2012.
\newblock \doi{10.1007/s10579-011-9145-0}.
\newblock URL \url{https://doi.org/10.1007/s10579-011-9145-0}.

\bibitem[Jaeger et~al.(2019)Jaeger, Trivedi, and Stadtschnitzer]{mdvrsource}
Hagen Jaeger, Dhaval Trivedi, and Michael Stadtschnitzer.
\newblock {Mobile Device Voice Recordings at King's College London (MDVR-KCL)
  from both early and advanced Parkinson's disease patients and healthy
  controls}, 2019.
\newblock URL \url{https://zenodo.org/records/2867216}.

\bibitem[Verde and Sannino(2018)]{voicedsource}
Laura Verde and Giovanna Sannino.
\newblock {VOICED} database.
\newblock \emph{PhysioNet}, 2018.
\newblock \doi{10.13026/C25Q2N}.
\newblock URL \url{https://physionet.org/content/voiced/1.0.0/}.
\newblock Version 1.0.0.

\bibitem[Lea et~al.(2021)Lea, Mitra, Joshi, Kajarekar, and
  Bigham]{sep28ksource}
Colin Lea, Vikramjit Mitra, Aparna Joshi, Sachin Kajarekar, and Jeffrey~P.
  Bigham.
\newblock {SEP-28k: A Dataset for Stuttering Event Detection from Podcasts with
  People Who Stutter}.
\newblock In \emph{ICASSP 2021 - 2021 IEEE International Conference on
  Acoustics, Speech and Signal Processing (ICASSP)}, pages 6798--6802, 2021.
\newblock \doi{10.1109/icassp39728.2021.9413520}.
\newblock URL \url{https://doi.org/10.1109/icassp39728.2021.9413520}.

\bibitem[Piczak(2015)]{esc50source}
Karol~J. Piczak.
\newblock {ESC: Dataset for Environmental Sound Classification}.
\newblock In \emph{Proceedings of the 23rd ACM international conference on
  Multimedia}, pages 1015--1018, 2015.
\newblock \doi{10.1145/2733373.2806390}.
\newblock URL \url{https://doi.org/10.1145/2733373.2806390}.

\bibitem[Cao et~al.(2014)Cao, Cooper, Keutmann, Gur, Nenkova, and
  Verma]{cremadsource}
Houwei Cao, David~G. Cooper, Michael~K. Keutmann, Ruben~C. Gur, Ani Nenkova,
  and Ragini Verma.
\newblock {CREMA-D: Crowd-Sourced Emotional Multimodal Actors Dataset}.
\newblock \emph{IEEE Transactions on Affective Computing}, 5\penalty0
  (4):\penalty0 377--390, 2014.
\newblock \doi{10.1109/taffc.2014.2336244}.
\newblock URL \url{https://doi.org/10.1109/taffc.2014.2336244}.

\bibitem[Livingstone and Russo(2018)]{ravdesssource}
Steven~R. Livingstone and Frank~A. Russo.
\newblock {The Ryerson Audio-Visual Database of Emotional Speech and Song
  (RAVDESS): A dynamic, multimodal set of facial and vocal expressions in North
  American English}.
\newblock \emph{PLOS ONE}, 13\penalty0 (5):\penalty0 e0196391, 2018.
\newblock \doi{10.1371/journal.pone.0196391}.
\newblock URL \url{https://doi.org/10.1371/journal.pone.0196391}.

\bibitem[Gratch et~al.(2014)Gratch, Artstein, Lucas, Stratou, Scherer,
  Nazarian, Wood, Boberg, DeVault, Marsella, Traum, Rizzo, and
  Morency]{daicsource}
Jonathan Gratch, Ron Artstein, Gale Lucas, Giota Stratou, Stefan Scherer,
  Angela Nazarian, Rachel Wood, Jill Boberg, David DeVault, Stacy Marsella,
  David Traum, Skip Rizzo, and Louis-Philippe Morency.
\newblock The distress analysis interview corpus of human and computer
  interviews.
\newblock In \emph{Proceedings of the Ninth International Conference on
  Language Resources and Evaluation ({LREC}'14)}, pages 3123--3128, 2014.
\newblock URL \url{https://aclanthology.org/L14-1421/}.

\bibitem[Shen et~al.(2022)Shen, Yang, and Lin]{eatdsource}
Ying Shen, Huiyu Yang, and Lin Lin.
\newblock {Automatic Depression Detection: an Emotional Audio-Textual Corpus
  and A GRU/BiLSTM-Based Model}.
\newblock In \emph{ICASSP 2022 - 2022 IEEE International Conference on
  Acoustics, Speech and Signal Processing (ICASSP)}, pages 6247--6251, 2022.
\newblock \doi{10.1109/icassp43922.2022.9746569}.
\newblock URL \url{https://doi.org/10.1109/icassp43922.2022.9746569}.

\bibitem[Jin et~al.(2020)Jin, Pan, Oufattole, Weng, Fang, and Szolovits]{medqa}
Di~Jin, Eileen Pan, Nassim Oufattole, Wei-Hung Weng, Hanyi Fang, and Peter
  Szolovits.
\newblock What disease does this patient have? a large-scale open domain
  question answering dataset from medical exams.
\newblock \emph{arXiv preprint arXiv:2009.13081}, 2020.
\newblock URL \url{https://arxiv.org/abs/2009.13081}.

\bibitem[Vilares and G{\'o}mez-Rodr{\'i}guez(2019)]{headqasource}
David Vilares and Carlos G{\'o}mez-Rodr{\'i}guez.
\newblock {HEAD}-{QA}: A healthcare dataset for complex reasoning.
\newblock In \emph{Proceedings of the 57th Annual Meeting of the Association
  for Computational Linguistics}, pages 960--966, 2019.
\newblock \doi{10.18653/v1/P19-1092}.
\newblock URL \url{https://aclanthology.org/P19-1092/}.

\bibitem[Chen et~al.(2025{\natexlab{b}})Chen, Fang, Singla, and
  Dredze]{medbulletssource}
Hanjie Chen, Zhouxiang Fang, Yash Singla, and Mark Dredze.
\newblock Benchmarking large language models on answering and explaining
  challenging medical questions.
\newblock In \emph{Proceedings of the 2025 Conference of the Nations of the
  Americas Chapter of the Association for Computational Linguistics: Human
  Language Technologies (Volume 1: Long Papers)}, pages 3563--3599,
  2025{\natexlab{b}}.
\newblock \doi{10.18653/v1/2025.naacl-long.182}.
\newblock URL \url{https://aclanthology.org/2025.naacl-long.182/}.

\bibitem[Bensoussan et~al.(2025)Bensoussan, Sigaras, Rameau, Elemento, Powell,
  Dorr, Payne, Ravitsky, B\'elisle-Pipon, Bahr, et~al.]{bridge2ai2025v3}
Yael Bensoussan, Alexandros Sigaras, Anais Rameau, Olivier Elemento, Maria
  Powell, David Dorr, Philip Payne, Vardit Ravitsky, Jean-Christophe
  B\'elisle-Pipon, Ruth Bahr, et~al.
\newblock {Bridge2AI-Voice}: An ethically-sourced, diverse voice dataset linked
  to health information.
\newblock \emph{PhysioNet}, 2025.
\newblock \doi{10.13026/k81f-qr68}.
\newblock URL \url{https://physionet.org/content/b2ai-voice/3.0.0/}.
\newblock Version 3.0.0.

\end{thebibliography}
}

\clearpage
\appendix
\setcounter{topnumber}{2}
\setlength{\textfloatsep}{10pt plus 2pt minus 2pt}
\setlength{\floatsep}{8pt plus 2pt minus 2pt}
\setlength{\intextsep}{8pt plus 2pt minus 2pt}

\section{Data Synthesis Details}
\label{app:data_synthesis_details}

\subsection{Department--Scenario Design}
\label{app:department_scenario_design}

The audio-conditioned medical dialogue dataset is synthesized from a controlled
department--scenario design. We define 25 medical departments and 15 clinical
interaction scenarios, yielding 375 department--scenario combinations. This
design allows the generated dialogues to cover diverse specialties, user roles,
clinical environments, and urgency levels.

\begin{table}[!htbp]
\centering
\begin{minipage}[t]{0.55\linewidth}
\vspace{0pt}
\caption{Medical departments used for synthesis.}
\label{tab:synthesis_departments}
\footnotesize
\setlength{\tabcolsep}{3pt}
\begin{tabularx}{\linewidth}{@{}>{\raggedright\arraybackslash}p{0.24\linewidth}>{\raggedright\arraybackslash}X@{}}
\toprule
\textbf{Group} & \textbf{Departments} \\
\midrule
Internal medicine
& Neurology; Cardiology; Respiratory Medicine; Gastroenterology;
Endocrinology; Nephrology; Hematology; Rheumatology and Immunology;
Infectious Disease; Medical Oncology. \\
Surgery
& General Surgery; Orthopedics; Neurosurgery; Cardiothoracic Surgery;
Urology. \\
Women and children
& Gynecology; Obstetrics; Pediatrics. \\
Specialty care
& Ophthalmology; ENT; Dermatology; Psychiatry; Rehabilitation Medicine. \\
Acute and critical care
& Emergency Medicine; ICU. \\
\bottomrule
\end{tabularx}
\end{minipage}\hfill
\begin{minipage}[t]{0.42\linewidth}
\vspace{0pt}
\caption{Controlled synthesis space.}
\label{tab:synthesis_space_summary}
\footnotesize
\setlength{\tabcolsep}{3pt}
\begin{tabularx}{\linewidth}{@{}Xr@{}}
\toprule
\textbf{Item} & \textbf{Value} \\
\midrule
Medical departments & 25 \\
Interaction scenarios & 15 \\
Department--scenario pairs & 375 \\
Topics per pair & 50 \\
Full topic pool & 18,750 \\
Sampled topics for dialogue synthesis & 5,000 \\
Samples per pair (approx.) & 13--14 \\
\bottomrule
\end{tabularx}
\end{minipage}
\end{table}

\begin{table}[!htbp]
\centering
\caption{Clinical interaction scenarios used for dialogue synthesis.}
\label{tab:synthesis_scenarios}
\footnotesize
\renewcommand{\arraystretch}{0.95}
\begin{tabularx}{\textwidth}{r>{\raggedright\arraybackslash}p{0.22\textwidth}X}
\toprule
\textbf{ID} & \textbf{User role} & \textbf{Scenario} \\
\midrule
1
& Doctor
& Doctor and patient conversing in a consultation room, where the doctor seeks
medical AI assistance midway. \\
2
& Patient
& Patient at home in a bedroom, directly having a voice consultation dialogue
with medical AI. \\
3
& Nurse
& Nurse at an emergency triage desk receiving patients and consulting a medical
AI system. \\
4
& Elderly patient
& Elderly person at a community health center communicating with medical AI
through a device. \\
5
& Paramedic
& Ambulance doctor in a moving vehicle urgently seeking medical AI support. \\
6
& Pharmacist
& Pharmacist at a hospital pharmacy window consulting medical AI when
encountering medication-related questions. \\
7
& Researcher
& Researcher in a university office having a medical professional dialogue with
medical AI. \\
8
& Family member
& Patient family member outside an ICU ward confiding in and consulting medical
AI. \\
9
& Resident doctor
& Resident trainee doing ward rounds in an inpatient department and asking
medical AI via tablet. \\
10
& Surgeon
& Surgeon in a sterile operating room interacting with medical AI through voice. \\
11
& Rehabilitation therapist
& Rehabilitation therapist in a physical therapy room activating medical AI for
guidance assistance. \\
12
& Pregnant woman
& Pregnant woman at home in a living room communicating with medical AI via
smartphone. \\
13
& Laboratory doctor
& Laboratory doctor in a hospital lab seeking medical AI opinion when facing
test reports. \\
14
& Dietitian
& Dietitian in a hospital meal preparation room seeking medical AI advice. \\
15
& Outpatient visitor
& Patient at a hospital outpatient lobby talking with medical AI through a
guidance kiosk. \\
\bottomrule
\end{tabularx}
\end{table}

\section{Data Statistics and Filtering}
\label{app:data_details}

\begin{table}[htbp]
\centering
\begin{minipage}[t]{0.49\linewidth}
\vspace{0pt}
\caption{Dialogue rubrics before filtering.}
\label{tab:rubric_data_statistics}
\small
\setlength{\tabcolsep}{3pt}
\begin{tabularx}{\linewidth}{@{}Xr@{}}
\toprule
\textbf{Statistic} & \textbf{Value} \\
\midrule
Dialogue turns with rubrics & 8,000 \\
Unique conversations & 2,000 \\
Mean number of positive items & 6.32 \\
Range of positive items & 5--10 \\
Mean number of negative items & 3.30 \\
Range of negative items & 2--5 \\
\bottomrule
\end{tabularx}
\end{minipage}\hfill
\begin{minipage}[t]{0.49\linewidth}
\vspace{0pt}
\caption{Baseline-based filtering statistics.}
\label{tab:baseline_filtering}
\small
\setlength{\tabcolsep}{3pt}
\begin{tabularx}{\linewidth}{@{}Xr@{}}
\toprule
\textbf{Statistic} & \textbf{Value} \\
\midrule
Input samples & 8,000 \\
Retained samples & 7,127 \\
Removed high-reward samples & 556 \\
Removed non-positive-reward samples & 317 \\
Samples with missing scores & 0 \\
\bottomrule
\end{tabularx}
\end{minipage}
\end{table}

Filtering retains examples with baseline reward strictly between 0 and 0.9.

\section{Reward and Severity Details}
\label{app:reward_details}

When severity labels are explicitly available, we use them directly. Otherwise,
severity is inferred from item weights and safety-sensitive medical keywords.
Negative items with weight magnitude $|w_j|\geq 8$ or containing high-risk concepts such as
contraindication, unsafe dosage, stroke, sepsis, anaphylaxis, airway compromise,
respiratory failure, bleeding, overdose, urgent escalation, or emergency care
are treated as critical. Otherwise, negative items with weight magnitude
$|w_j|\geq 5$ are treated as major. The
remaining negative items are treated as minor. 

\noindent  \textbf{Length and truncation penalties.}
For response length $L$ in policy-tokenizer tokens, the length penalty increases
linearly from zero at $f$ to $p_{\max}$ at $h$:
\begin{equation}
\begin{aligned}
P_{\mathrm{len}}(y)
&= p_{\max}\min\!\left(
    1,\max\!\left(0,\frac{L-f}{h-f}\right)
   \right), \\
P_{\mathrm{trunc}}(y)
&= p_{\mathrm{EOS}}\,
   \mathbf{1}\!\left\{
   y\text{ is nonempty and lacks EOS}
   \right\}.
\end{aligned}
\end{equation}
Medical dialogue RL and audio-integrated dialogue RL use $p_{\max}=p_{\mathrm{EOS}}=0.15$ and a 256-token completion budget.
The free/hard thresholds are 1.10/1.50 times the reference length, rounded up
and adjusted to satisfy $1\leq f<h\leq256$; without a reference, they are
120/200 tokens. EOS denotes \texttt{<|im\_end|>} and is excluded from length
counting.

\section{Training Configuration}
\label{app:training_details}

\noindent  \textbf{Training setup.}
We initialize from Qwen2.5-Omni-7B and sequentially perform medical SFT,
medical dialogue RL and audio-integrated dialogue RL.
Table~\ref{tab:training_configuration} summarizes the data sizes and
hyperparameters. 

\begin{table}[!htbp]
\centering
\caption{Training data sizes and key hyperparameters for SFT and the two RL stages.}
\label{tab:training_configuration}
\small
\begin{tabular}{lccc}
\toprule
Setting & SFT & \shortstack{Medical dialogue\\RL} & \shortstack{Audio-integrated\\dialogue RL} \\
\midrule
Training examples & 18,657 & 6,927 & 4,546 \\
Learning rate & $8\times10^{-6}$ & $5\times10^{-6}$ & $5\times10^{-7}$ \\
Per-device batch size & 8 & 4 & 4 \\
Completions per prompt & -- & 10 & 10 \\
\bottomrule
\end{tabular}
\end{table}

All stages use fused AdamW, BF16, gradient checkpointing, and seed 42.
SFT uses a cosine learning-rate schedule, while both RL stages use constant
learning rates after warmup. RL uses DAPO with asymmetric clipping thresholds
of 0.20/0.28 and KL regularization against each stage's initial policy.
SFT freezes the visual encoder and trains the language model and aligner;
the visual branch receives no image or video inputs during RL.

\subsection{Local judge training and validation}
\label{app:local_judge_details}
Both judges fine-tune \judgeModel{} for one epoch using LoRA
(rank 32, $\alpha=64$, dropout 0.05) and GPT-5.4 binary labels for individual
response--criterion pairs. 

To include responses spanning a range of quality levels, the dialogue data contain responses from GPT-4.1, GPT-4o-mini, and GPT-4.1-mini, with disjoint training and validation conversations. Audio training combines
current-policy responses with dialogue replay; audio splits are
disjoint by source-audio family, and replay uses only dialogue training data.
Audio evidence is supplied to the judges as reference text.

\begin{table}[!htbp]
\centering
\caption{Local judge data sizes and agreement with held-out GPT-5.4 labels.
Counts are response--criterion pairs; metrics are percentages.}
\label{tab:local_judge_validation}
\small
\begin{tabular}{lrr}
\toprule
Setting / metric & Dialogue judge & Audio judge \\
\midrule
Training examples & 112,514 & 22,872 \\
Validation examples & 2,900 & 3,771 \\
Parse success & 100.00 & 100.00 \\
Overall accuracy & 94.76 & 89.90 \\
Positive-item accuracy & 92.37 & 92.31 \\
Negative-item accuracy & 99.39 & 85.77 \\
\bottomrule
\end{tabular}
\end{table}

Baselines include seven general-purpose open-weight models:
Qwen2.5-Omni-7B~\citep{qwen25omni}, \href{https://huggingface.co/openbmb/MiniCPM-o-2_6}{MiniCPM-o-2.6}~\citep{minicpmv},
Baichuan-Omni-1.5~\citep{baichuanomni}, GLM-4-Voice-9B~\citep{glm4voice},
Step-Audio-2-mini~\citep{stepaudio2}, LLaMA-3.1-8B-Omni~\citep{llamaomni}, and
SpeechGPT-2.0-preview~\citep{speechgpt2}; closed-source references are
Gemini-2.5-Flash, Gemini-3.7-Flash, and GPT-Audio (2025-08-28).

\noindent  \textbf{ACI-Bench adaptation.}
The original ACI-Bench task generates visit notes from encounter transcripts~\citep{acibench}.
We instead extract patient--doctor exchanges and hold out a doctor utterance
as the reference response. The preceding patient utterance is converted to
speech, and earlier turns are supplied as text. The target is the original
doctor utterance from the transcript, not the visit note. Consequently, our
rubric scores measure dialogue-response quality under this adaptation and
are not the original ACI-Bench note-generation scores.

\noindent  \textbf{Evaluation inputs.}
Table~\ref{tab:evaluation_input_content} summarizes the inputs available to
each policy. Compared models receive matched task content in their respective
input formats.
ASR tasks additionally include a transcription instruction. MedVoice-Dial
provides user speech with its original TTS source text and reference-assistant
history.
\begin{table}[!htbp]
\centering
\caption{Policy-visible inputs within each evaluation setting. Compared models
receive the same task content, encoded using their respective model/API
formats. Current-target references and scoring rubrics remain judge-only.}
\label{tab:evaluation_input_content}
\footnotesize
\begin{tabularx}{\linewidth}{@{}>{\raggedright\arraybackslash}p{0.24\linewidth}*{3}{>{\raggedright\arraybackslash}X}@{}}
\toprule
Benchmark & Audio & Current-turn transcript & Supplied history \\
\midrule
LiveMedBench Clean-TTS & Spoken query & None & None \\
MMedFD-Lite & Spoken query & None & None \\
MedVoice-AID & Composed audio request & None & None \\
MedVoice-Dial & Current and preceding user speech & Original TTS source text & User audio and source text; reference-assistant text \\
ACI-Bench & Final patient turn & None & Text dialogue \\
HealthDial & Current user turn & None & Text dialogue \\
ThReadMed-QA & Current user turn & None & Text dialogue \\
LibriSpeech & Speech audio & None & None \\
AISHELL-1 & Speech audio & None & None \\
VoiceBench & Spoken query & None & None \\
OpenHermes-Audio & Spoken query & None & None \\
WildSpeech-Bench & Spoken query & None & None \\
\bottomrule
\end{tabularx}
\end{table}

\section{Audio Data}
\label{app:audio_details}

Table~\ref{tab:sft_source_composition} gives the three-source composition of
the final SFT training manifest; the 6,000 audio-integrated examples comprise
two 3,000-example sources. The 1,400 validation examples are counted separately.

\begin{table}[!htbp]
\centering
\caption{Composition of the final multimodal medical SFT training manifest
following selection and assembly. Counts are training examples.}
\label{tab:sft_source_composition}
\begin{tabular}{lr}
\toprule
\textbf{Supervision source} & \textbf{Training examples} \\
\midrule
Collected medical audio & 7,569 \\
Synthesized spoken dialogue & 5,088 \\
Audio-integrated dialogue & 6,000 \\
\midrule
Total & 18,657 \\
\bottomrule
\end{tabular}
\end{table}

\noindent  \textbf{Fine-grained composition and source references.}
Table~\ref{tab:medical_audio_categories} partitions the 7,569 selected training
examples in the collected medical-audio stream by source domain.
Table~\ref{tab:medical_audio_sources} lists the corresponding source references
and selected counts.

\begin{table}[!htbp]
\centering
\caption{Source-domain breakdown of the collected medical-audio SFT stream
in the final training manifest.}
\label{tab:medical_audio_categories}
\small
\begin{tabularx}{\linewidth}{@{}Xr@{}}
\toprule
Source domain & Training examples \\
\midrule
Lung and respiratory sounds & 2,697 \\
Pediatric respiratory sounds & 386 \\
Heart sounds & 365 \\
Cough and respiratory-health audio & 590 \\
Environmental health-related sounds & 34 \\
Voice and motor-speech disorders & 414 \\
Stuttering & 199 \\
Clinical voice features reconstructed as audio & 349 \\
Emotional speech & 339 \\
Depression-related speech & 100 \\
Pain-related vocalizations & 167 \\
Infant cries & 88 \\
Medical speech and transcription & 738 \\
Spoken medical intent & 163 \\
Medical QA converted to speech & 940 \\
\midrule
Total & 7,569 \\
\bottomrule
\end{tabularx}
\end{table}

\begin{table}[!htbp]
\centering
\caption{Audio-data provenance and references. Each linked dataset name points
to its source repository or dataset record. Counts are selected SFT training
examples; they do not represent the full size of the source datasets.}
\label{tab:medical_audio_sources}
\footnotesize
\begin{tabularx}{\linewidth}{@{}XrXr@{}}
\toprule
Source (linked reference) & Count & Source (linked reference) & Count \\
\midrule
\href{https://gitlab.com/techsupportHF/HF_Lung_V1}{HF Lung V1} & 2,575 & \href{https://github.com/CheyneyComputerScience/CREMA-D}{CREMA-D} & 200 \\
\href{https://bhichallenge.med.auth.gr/}{ICBHI 2017} & 122 & \href{https://zenodo.org/records/1188976}{RAVDESS} & 139 \\
\href{https://github.com/SJTU-YONGFU-RESEARCH-GRP/SPRSound}{SPRSound} & 386 & \href{https://dcapswoz.ict.usc.edu/}{DAIC-WOZ} & 27 \\
\href{https://physionet.org/content/challenge-2016/1.0.0/}{PhysioNet/CinC 2016} & 202 & \href{https://github.com/speechandlanguageprocessing/ICASSP2022-Depression}{EATD-Corpus} & 73 \\
\href{https://physionet.org/content/circor-heart-sound/1.0.3/}{CirCor DigiScope} & 163 & \href{https://physionet.org/content/tame-pain/1.0.0/}{TAME Pain} & 167 \\
\href{https://github.com/iiscleap/Coswara-Data}{Coswara} & 146 & \href{https://github.com/gveres/donateacry-corpus}{Donate-a-Cry} & 88 \\
\href{https://zenodo.org/records/4498364}{COUGHVID} & 444 & \href{https://huggingface.co/datasets/united-we-care/United-Syn-Med}{United-Syn-Med} & 476 \\
\href{https://github.com/karolpiczak/ESC-50}{ESC-50} & 34 & \href{https://huggingface.co/datasets/Hani89/medical_asr_recording_dataset}{Medical ASR Recording} & 99 \\
\href{https://www.cs.toronto.edu/~complingweb/data/TORGO/torgo.html}{TORGO} & 172 & \href{https://huggingface.co/datasets/aline-gassenn/MedDialog-Audio}{MedDialog-Audio} & 79 \\
\href{https://www.synapse.org/Synapse:syn53009474}{VOC-ALS} & 90 & \href{https://huggingface.co/datasets/sahilkadge/medical_audio_dataset}{Medical Audio Dataset} & 3 \\
\href{https://zenodo.org/records/2867216}{MDVR-KCL} & 7 & \href{https://huggingface.co/datasets/yassino3ss/youtube_medical_audio_dataset}{YouTube Medical Audio} & 81 \\
\href{https://huggingface.co/datasets/b-brave/speech_disorders_voice}{Speech Disorders Voice} & 110 & \href{https://huggingface.co/datasets/shreyas1104/medical-intent-audio-dataset}{Medical Intent Audio} & 163 \\
\href{https://physionet.org/content/voiced/1.0.0/}{VOICED} & 35 & \href{https://github.com/jind11/MedQA}{MedQA} & 715 \\
\href{https://github.com/apple-aiml-research/ml-stuttering-events-dataset}{SEP-28k} & 199 & \href{https://github.com/aghie/head-qa}{HEAD-QA} & 151 \\
\href{https://physionet.org/content/b2ai-voice/3.0.0/}{Bridge2AI-Voice (reconstructed)} & 349 & \href{https://github.com/HanjieChen/ChallengeClinicalQA}{MedBullets} & 74 \\
\bottomrule
\end{tabularx}
\end{table}

\FloatBarrier

\noindent  \textbf{Dataset references.}
Published descriptions accompany the respiratory sources HF Lung V1~\citep{hf_lungsource},
ICBHI~\citep{rocha2019respiratory}, SPRSound~\citep{sprsoundsource},
Coswara~\citep{coswarasource}, and COUGHVID~\citep{coughvidsource};
the cardiac sources PhysioNet/CinC 2016~\citep{physionet2016source} and
CirCor DigiScope~\citep{oliveira2022circor}; and the speech and voice sources
TORGO~\citep{torgosource}, MDVR-KCL~\citep{mdvrsource}, VOICED~\citep{voicedsource},
and SEP-28k~\citep{sep28ksource}.
Additional source descriptions cover ESC-50~\citep{esc50source},
CREMA-D~\citep{cremadsource}, RAVDESS~\citep{ravdesssource},
the DAIC interview corpus underlying DAIC-WOZ~\citep{daicsource},
EATD-Corpus~\citep{eatdsource}, MedQA~\citep{medqa}, HEAD-QA~\citep{headqasource},
and MedBullets~\citep{medbulletssource}.
The links in Table~\ref{tab:medical_audio_sources} identify the dataset records
and community releases used in this study.

\noindent  \textbf{Reconstructed Bridge2AI-Voice audio.}
\label{app:bridge2ai_reconstruction}
The 349 Bridge2AI-Voice training examples use waveforms reconstructed from
the spectrogram features in the v3.0.0 release~\citep{bridge2ai2025v3}.
These are derived audio, not the original clinical recordings, which are
not included in that PhysioNet release. The preprocessing pipeline uses
Griffin--Lim reconstruction and audio post-processing. Reconstruction may
alter clinically relevant acoustic details, so these examples should not
be interpreted as evidence of diagnostic performance on original
Bridge2AI-Voice recordings.

\section{Human Evaluation Details}
\label{app:human_evaluation}

\noindent  \textbf{Cohort and assignment.}
The physician evaluation compares \modelname{} with \baseModel{},
MiniCPM-o-2.6, Baichuan-Omni-1.5, GLM-4-Voice-9B, Step-Audio-2-mini,
and LLaMA-3.1-8B-Omni. The 30 cases cover respiratory medicine, pediatrics,
cardiology, neurology, and otorhinolaryngology. Each of the six physicians
completed 120 comparisons, comprising 18 comparisons shared by all reviewers
and 102 additional assignments unique to that reviewer. This yields 630
unique comparisons and 720 judgments. Twelve judgments were unjudgeable;
the remaining 708 judgments cover 621 comparisons with at least one valid
vote. The interface omitted model identifiers and randomized question order
and response position. Each physician spent approximately 1.5 hours completing the evaluation
and received RMB 1,000 in compensation.

\noindent  \textbf{Preference aggregation and uncertainty.}
We use the final submitted judgments and exclude unjudgeable votes.
For each comparison, a win, tie, and loss receive credits of 1, 0.5, and 0,
respectively. Direct pairwise preference credits average these values across
comparisons after averaging valid votes within each comparison.
Valid reviewer credits are averaged before fitting the
Bradley--Terry model, so a shared comparison does not outweigh a comparison
assigned to one reviewer. Scores use the same logistic fit and scale as the
automated Arena analysis, with \baseModel{} anchored at 1000. We resample
source-case clusters 2,000 times to obtain percentile intervals, retaining
all comparisons from a sampled case together.

\noindent  \textbf{Shared-item agreement.}
Responses are aligned by model identity rather than displayed A/B position;
ties form a separate category and unjudgeable votes are treated as missing.
The 18 shared comparisons contain 105 valid judgments. There are 152 agreeing
pairs among 255 comparable reviewer pairs, giving 59.6\% agreement
(95\% item-bootstrap interval: 46.0--73.8\%). Four comparisons receive the same
choice from all six physicians. Agreement is 52.9\% between the two attending
physicians, 59.8\% among resident physicians, and 60.3\% across the two groups.

\section{Prompt Templates and Evaluation Protocols}
\label{app:prompt_templates}

The prompt templates in this section make the data construction and evaluation
protocols reproducible. Double braces denote fields substituted from the locked
manifest. The code blocks show the core templates used in the reported runs;
minor dataset-specific metadata is passed through the same JSON fields. 

\begin{table}[!htbp]
\centering
\caption{Prompt inventory for data construction, training, and evaluation.}
\label{tab:prompt_inventory}
\footnotesize
\setlength{\tabcolsep}{3pt}
\begin{tabularx}{\linewidth}{@{}l>{\raggedright\arraybackslash}p{0.28\linewidth}X@{}}
\toprule
\textbf{ID} & \textbf{Use} & \textbf{Output contract} \\
\midrule
P1 & Topic generation & One concise clinical topic per department--scenario pair \\
P2 & Spoken dialogue synthesis & JSON alternating user and assistant turns \\
P3 & Audio-integrated case synthesis & JSON case, audio script, reference, and signed rubric \\
P4 & Policy input & Multimodal messages with audio markers and transcript fields \\
P5 & Rubric generation and repair & Instance-specific signed rubric JSON \\
P6 & Local item judge & Binary labels for every positive and negative item \\
P7 & Medical benchmark scoring & Criterion-level JSON or official G-Eval JSON \\
P8 & General speech scoring & ASR text, IFEval score, or GPT-5.5 scalar score \\
\bottomrule
\end{tabularx}
\end{table}

\subsection{Data synthesis prompts}
\label{app:prompt_data_synthesis}

\noindent  \textbf{Topic and spoken-dialogue synthesis.}
Topic generation receives one controlled department--scenario pair. Dialogue
synthesis then receives the selected topic and the same scenario metadata. Both
stages require natural spoken English, role-consistent language, and structured
JSON output.

\begin{promptcode}
[system]
You are a senior clinical data curator and medical dialogue writer.
Return exactly one JSON object and no markdown. Keep all user-facing
utterances natural, clinically grounded, and suitable for text-to-speech.
[input]
department: {{department}}
scenario: {{scenario}}
user_role: {{user_role}}
topic: {{topic}}
[requirements]
Generate an alternating user--assistant conversation with four user turns
and four assistant turns. Match the vocabulary and authority to the role,
make the turns progress logically, avoid bullet points and annotation language,
and do not reveal hidden labels, filenames, or grading criteria.
[output]
{"dialogues":[{"role":"user","content":"..."},
{"role":"assistant","content":"..."}]}
\end{promptcode}

\noindent  \textbf{Audio-integrated case synthesis.}
The audio case builder receives the source-audio analysis and must separate
observable acoustic evidence from hidden source metadata. It produces the
spoken request, reference response, audio construction instructions, and
instance-specific rubric in one JSON object.

\begin{promptcode}
[system]
You are a senior clinical data curator creating single-turn medical-audio data.
Return exactly one JSON object and no markdown. Ground every statement in the
supplied audio evidence; never infer a diagnosis from a recording alone.
[input]
source_modality: {{source_modality}}
source_analysis: {{source_analysis}}
department: {{department}}
scenario: {{scenario}}
user_role: {{user_role}}
[requirements]
Create one natural current user decision, an audio construction script, a
reference assistant response, and a signed rubric. State recording/device
context when relevant, preserve the real source recording, use proportionate
safety guidance, and keep unsupported diagnoses and metadata out of speech.
\end{promptcode}

\subsection{Training input and reward prompts}
\label{app:prompt_training}

The canonical policy system message is shared by the medical dialogue SFT and
RL inputs. Transcript availability in evaluation depends on the benchmark
(Table~\ref{tab:evaluation_input_content}). The current target reference
response, rubric, and reward are not visible to the policy during rollout;
preceding reference-assistant turns form the supplied dialogue history.

\begin{promptcode}
[system]
You are a professional medical voice AI assistant. Your role is to:
- Provide accurate, empathetic medical information and support.
- Help users understand symptoms, treatments, and medical conditions.
- Give medication, post-operative, chronic-care, and emergency guidance.
- Explain medical information and test results clearly.
Always remain professional, supportive, and patient-centered.
[optional single-turn audio suffix]
Answer the current question first, use only supported audible evidence, match
the user's role, and give the next step and uncertainty when relevant.
[user]
<audio>
{{current_user_transcript}}
[assistant history]
{{causal_dialogue_history}}
\end{promptcode}

\noindent  \textbf{Instance-specific rubric generation.}
The rubric writer sees the full preceding conversation, the final request, and the
reference answer. It returns 8--12 atomic binary criteria, with positive
items in the majority and signed weights reflecting clinical importance.

\begin{promptcode}
[system]
You are a senior clinical rubric writer designing reward criteria for
medical-AI reinforcement learning. Create a self-contained rubric for a new
candidate response to the final user request.
[requirements]
Use the complete dialogue history. Reward specific, clinically grounded,
useful content; detect hallucinations, unsafe advice, false reassurance,
contradictions, and dangerous overclaims. Do not require reference wording.
Every item must be independently judgeable with a binary answer. Use positive
weights +1..+10 and negative weights -1..-10; reserve magnitude 8..10 for
clinically consequential decisions.
[output]
{"rubric_version":"instance-signed-v1",
 "clinical_focus":"...",
 "historical_constraints":["..."],
 "positive_items":[{"id":"P1","scope":"current_turn",
   "criterion":"Does the response ...?","weight":8,"rationale":"..."}],
 "negative_items":[{"id":"N1","scope":"dialogue_consistency",
   "criterion":"Does the response incorrectly ...?","weight":-8,
   "rationale":"..."}]}
\end{promptcode}

\noindent  \textbf{Rubric review and repair.}
A second reviewer prompt removes redundant or logically inverted items before
the rubric enters the training manifest.

\begin{promptcode}
[system]
You are a strict reviewer of a signed medical RL rubric. Repair the candidate
rubric while preserving its schema and clinical intent. Merge redundant items,
remove negative items that only invert a positive item, keep omission penalties
only when the omission creates immediate safety risk, preserve supported
historical constraints, keep 8--12 total items with positive items in the
majority, and renumber IDs consecutively. Return only repaired JSON.
[user]
{{candidate_rubric}}
\end{promptcode}

\noindent  \textbf{Local item-level judge.}
The local \judgeModel{} LoRA judge receives one candidate response and
one rubric criterion per query. It returns \texttt{\{"answer": 0\}} or
\texttt{\{"answer": 1\}}; a label of 1 indicates that the response satisfies a positive criterion
or exhibits the prohibited behavior described by a negative criterion.
The following compact template preserves this single-item interface.

\begin{promptcode}
[system]
Judge exactly one criterion using the causal dialogue and judge-only evidence.
Return only {"answer": 0} or {"answer": 1}. Answer 1 when the criterion is true
of the current candidate. Negative items require affirmative prohibited claims
or actions, not quotation, denial, warnings, or marking a claim unsupported.
Treat Gold Audio Evidence as authoritative; do not infer findings from an
<audio> placeholder or from the candidate response.
[user]
## Complete Dialogue Before Candidate Response
{{history}}
## Gold Audio Evidence (Judge-Only Reference)
{{audio_evidence_summary}}
## Text Constraints (Judge-Only Reference)
{{text_constraints}}
## Historical Constraints
{{historical_constraints}}
## Candidate Assistant Response
{{candidate}}
## Single Rubric Criterion
Scope: {{scope}}
Criterion: {{criterion}}
Return ONLY {"answer": 0} or {"answer": 1}.
\end{promptcode}

\subsection{Medical benchmark judge prompts}
\label{app:prompt_medical_evaluation}

The external medical scorers are separate from the local RL judges. LiveMedBench Clean-TTS
uses one criterion at a time, MMedFD-Lite uses the released batch G-Eval
format, and ACI-Bench, HealthDial, ThReadMed-QA, MedVoice-AID, and MedVoice-Dial
use signed instance-rubric scoring with GPT-5.4 at low reasoning effort.

\begin{promptcode}
[system]
You are a strict medical AI evaluator. Judge one candidate response against
one supplied criterion using only the patient narrative and current request.
For a negatively weighted criterion, mark it true only when the undesirable
behavior is present. Return JSON only.
[user]
Patient narrative: {{narrative}}
Patient request: {{request}}
Candidate response: {{candidate}}
Rubric criterion: {{criterion}}
{"explanation":"brief evidence-based reason",
 "criteria_met":true}
\end{promptcode}

\begin{promptcode}
[MMedFD-Lite G-Eval]
Given a user query and a model reply, evaluate the response quality. Identify
repetition, off-topic content, role confusion, contradictions, and unsafe advice.
Return one JSON object containing the item id, an integer score from 1 to 10,
a short rationale, and the binary failure flags.
\end{promptcode}

For signed instance-rubric evaluation, the judge also receives the preceding dialogue, constraints established in earlier
  turns, all positive and negative rubric items, and a reference response available only to the judge. Before computing the
  signed score, the scorer verifies that each rubric item has been assessed exactly once. Arena evaluation uses a separate
  blinded pairwise prompt and reports Bradley--Terry scores, with model identities and training stages hidden from the judge.

\begin{promptcode}
[Arena pairwise judge]
[system]
You are an independent blinded evaluator of medical assistant responses. Use
the causal history and current user audio. Prefer medically accurate, safe,
grounded, directly useful, and clear answers. Penalize invented patient facts,
unsupported audio findings, dangerous advice, and irrelevant content. Do not
favor any model or reward verbosity. Return JSON only.
[user]
Causal text history: {{history}}
Current user audio: {{audio}}
Candidate A: {{candidate_A}}
Candidate B: {{candidate_B}}
{"winner":"A|B|tie|unjudgeable", "both_bad":false,
 "decisive_dimension":"...", "reason":"...",
 "audio_understood":true}
\end{promptcode}

\subsection{General speech evaluation prompts}
\label{app:prompt_general_speech}

All models are evaluated on the same fixed examples using greedy decoding.
The prompt shown to the policy differs by benchmark, while the text query,
reference answer, or checklist used by a judge is kept outside the policy input.

\begin{promptcode}
[LibriSpeech ASR]
system: You are a helpful assistant.
user: <audio>
      Transcribe the spoken content exactly. Output only the transcription.
[AISHELL-1 ASR]
system: You are a careful audio understanding assistant.
user: <audio>
      Please accurately transcribe the Chinese speech in the audio and output only the transcription.
[VoiceBench-IFEval and VoiceBench-CommonEval]
system: You are a helpful assistant.
user: <audio>
\end{promptcode}

\begin{promptcode}
[VoiceBench-CommonEval judge]
system: You are a helpful assistant who tries to help answer the user's question.
user:
I need your help to evaluate the performance of several models in the speech
interaction scenario. The models receive a speech input from the user and
need to understand and respond to it. Rate the response using the user-input
transcription [Instruction] and the model-output transcription [Response].
Please evaluate the response on a scale of 1 to 5:
1 point: largely irrelevant, incorrect, or fails to address the query.
2 points: somewhat relevant but inaccurate or incomplete.
3 points: relevant and mostly accurate but not concise.
4 points: relevant, accurate, and concise.
5 points: exceptionally relevant, accurate, and to the point.
### [Instruction]: {{query}}
### [Response]: {{candidate}}
Output the score only without explanations.
[OpenHermes-Audio judge]
[Reference Answer]
{{reference}}
[Model Answer]
{{candidate}}
[Question]
{{question}}
Rate accuracy and relevance from 0 to 5. A refusal or completely irrelevant
answer receives 0; 3 means generally aligned but incomplete; 5 means highly
accurate, relevant, and detailed. Return:
Explanation: ...
Rating: <integer>
[WildSpeech-Bench judge]
User query: {{query}}
AI response: {{candidate}}
Checklist: {{checklist}}
Rate spoken-response quality from 1 to 10, considering instruction adherence,
correctness, relevance, repetition, and the characteristics of voice dialogue.
Return:
Strengths: ...
Weaknesses: ...
Score: <integer>
\end{promptcode}

GPT-5.5 evaluates CommonEval, OpenHermes-Audio, and WildSpeech-Bench with
three independent calls per response. The requests use a 1,024-token completion
limit, top-$p=0.95$, and temperatures 0.5, 0.7, and 0.5 in the same order as
those three benchmarks. ASR uses WER/CER after the fixed normalization rules;
paired intervals use 20,000 bootstrap replicates. 

\noindent  \textbf{Prompt provenance.}
The reported tables use the prompts above and the corresponding frozen script
versions.

\section{Complete Ablation Results}
\label{app:ablation_results}

Table~\ref{tab:rl_only_ablation} provides the complete numerical results for
Figure~\ref{fig:sft_source_ablation}.

\begin{table}[!htbp]
\centering
\caption{Ablation of multimodal medical supervised fine-tuning and its training
sources. SFT omits both RL stages. RL-only starts directly from
Qwen2.5-Omni-7B, while the source-removal
variants omit the named SFT source before applying the same medical dialogue RL
followed by audio-integrated dialogue RL. The full pipeline applies SFT followed
by these two RL stages. All values are GPT-5.4 signed instance-rubric scores under the
same evaluation protocol.}
\label{tab:rl_only_ablation}
\small
\setlength{\tabcolsep}{5pt}
\resizebox{\textwidth}{!}{
\begin{tabular}{l ccccc}
\toprule
\textbf{Training pipeline}
& \textbf{MedVoice-AID}
& \textbf{MedVoice-Dial}
& \textbf{ACI-Bench}
& \textbf{HealthDial}
& \textbf{ThReadMed-QA} \\
\midrule
Qwen2.5-Omni-7B
& 0.3341 & 0.3237 & 0.2002 & 0.4489 & 0.1800 \\
SFT
& 0.7666 & 0.4257 & 0.2168 & 0.4106 & 0.2034 \\
RL-only
& 0.3033 & 0.4198 & 0.1297 & 0.3489 & 0.1641 \\
w/o collected medical audio
& 0.7552 & 0.4743 & 0.1732 & 0.4665 & 0.2705 \\
w/o synthesized spoken dialogue
& 0.7505 & 0.4587 & 0.2181 & 0.4570 & 0.2096 \\
w/o audio-integrated dialogue
& 0.5957 & 0.4725 & 0.1836 & 0.4749 & 0.3058 \\
\modelname{}
& \textbf{0.7900} & \textbf{0.5225} & \textbf{0.2263}
& \textbf{0.5022} & \textbf{0.3403} \\
\bottomrule
\end{tabular}}
\end{table}

\end{document}